\documentclass{emulateapj}
\usepackage{epsfig,float,longtable}
\def\gap{\lower.5ex\hbox{$\; \buildrel > \over \sim \;$}}
\def\lap{\lower.5ex\hbox{$\; \buildrel < \over \sim \;$}}
\def\XMM{{\it XMM-Newton }}
\makeatletter

\def\fudge#1#2{%
\addtolength\textheight{#1}%
\@colroom\textheight
\vsize\textheight
\@colht\textheight
\def\LS@rot{%
  \setbox\@outputbox\vbox{\hbox{\kern-#2\rotatebox{90}{\box\@outputbox}}}}%
\clearpage}

\makeatother

\begin{document}
\sloppy
\title{A Deep {\it XMM-Newton} Survey of M33: Point Source Catalog, Source Detection and Characterization of Overlapping Fields}
\author{Benjamin F. Williams\altaffilmark{1},
Brian Wold\altaffilmark{1},
Frank Haberl\altaffilmark{2},
Kristen Garofali\altaffilmark{1},
William P. Blair\altaffilmark{3},
Terrance J. Gaetz\altaffilmark{4},
K. D. Kuntz\altaffilmark{3},
Knox S. Long\altaffilmark{5},
Thomas G. Pannuti\altaffilmark{6},
Wolfgang Pietsch\altaffilmark{2},
Paul P. Plucinsky\altaffilmark{4},
P. Frank Winkler\altaffilmark{7}
}

\altaffiltext{1}{University of Washington Astronomy Department, Box 351580, Seattle, WA, 98195; ben@astro.washington.edu;garofali@astro.washington.edu; chevron.seven@gmail.com; kgarofali@gmail.com}

\altaffiltext{2}{Max-Planck-Institut f\"ur extraterrestrische Physik, Giessenbachstra{\ss}e, 85748 Garching, Germany; fwh@mpe.mpg.de; wnp@mpe.mpg.de}

\altaffiltext{3}{Department of Physics and Astronomy, Johns Hopkins University, 3400 North Charles Street, Baltimore, MD 21218; kuntz@pha.jhu.edu; wpb@pha.jhu.edu}

\altaffiltext{4}{Harvard-Smithsonian Center for Astrophysics, 60
Garden Street, Cambridge, MA 02138; plucinsk@head.cfa.harvard.edu; gaetz@head.cfa.harvard.edu}

\altaffiltext{5}{Space Telescope Science Institute, 3700 San Martin Drive, Baltimore, MD 21218; long@stsci.edu}

\altaffiltext{6}{Space Science Center, Department of Earth and Space Sciences, 235 Martindale Drive, Morehead State University, Morehead, KY 40351; t.pannuti@moreheadstate.edu}

\altaffiltext{7}{Department of Physics, Middlebury College, Middlebury, VT 05753; winkler@middlebury.edu}

\keywords{X-rays: binaries --- galaxies: individual (M33) --- X-rays: stars}

\begin{abstract}

We have obtained a deep 8-field {\it XMM-Newton} mosaic of M33 covering the galaxy out to the D$_{25}$ isophote and beyond to a limiting 0.2--4.5 keV unabsorbed flux of 5$\times$10$^{-16}$ erg cm$^{-2}$ s$^{-1}$ (L${>}$4$\times$10$^{34}$ erg s$^{-1}$ at the distance of M33).  These data allow complete coverage of the galaxy with high sensitivity to soft sources such as diffuse hot gas and supernova remnants.  Here we describe the methods we used to identify and characterize 1296 point sources in the 8 fields.  We compare our resulting source catalog to the literature, note variable sources, construct hardness ratios, classify soft sources, analyze the source density profile, and measure the X-ray luminosity function.  As a result of the large effective area of \XMM below 1 keV, the survey contains many new soft X-ray sources.  The radial source density profile and X-ray luminosity function for the sources suggests that only $\sim$15\% of the 391 bright sources with L${>}$3.6$\times$10$^{35}$ erg s$^{-1}$ are likely to be associated with M33, and more than a third of these are known supernova remnants. The log(N)--log(S) distribution, when corrected for background contamination, is a relatively flat power-law with a differential index of 1.5, which suggests many of the other M33 sources may be high-mass X-ray binaries. Finally, we note the discovery of an interesting new transient X-ray source, which we are unable to classify.

\end{abstract}

\section{Introduction}

There are only two spiral galaxies nearby enough to resolve individual
X-ray binaries and supernova remnants with {\it XMM-Newton}: M31 and
M33.  M31 has already been well-observed, with overlapping 60~ks
observations covering the entire D$_{25}$ extent of the disk
\citep{stiele2011}.  However, at this point M33 has only been observed
with relatively short ($\sim$10~ks) exposures.  These have been taken
over a relatively long time baseline and cover the entire galaxy, but
have not allowed many detailed spectral studies or even a very deep
X-ray luminosity function (XLF) to be measured \citep{pietsch2004,misanovic2006}.  

M33, a late-type Sc spiral, is ideal for studying X-ray point source
populations because of its proximity \citep[817$\pm$58
  kpc;][]{FreedmanDistance} and its relatively low inclination
\citep[{\it i} = 54$^{\circ}$;][]{devaucouleurs1991}.  M33 also has
low foreground absorption \citep[N$_{H}\approx$6 x
  10$^{20}$cm$^{-2}$,][]{StarkAbsorption}, simplifying detection of
resident discrete X-ray sources as well as interpretation of the
properties of the sources.  Multi-wavelength surveys have revealed a
large X-ray source population in M33 with many interesting variables
(\citealp[e.g.,][]{long1981,markert1983,schulman1995,long1996,haberl2001,pietsch2003,pietsch2004}
[hereafter P04]; \citealp{grimm2005,misanovic2006} [hereafter M06];
\citealp{pietsch2006,grimm2007,plucinsky2008,williams2008,Tullmann}
        [hereafter T11]), a rich supernova remnant (SNR) population
        \citep[137 total; 55 candidates, 82
          confirmed;][]{dodorico1980,long1990,gordon1998,LongSNR}, and
        significant hot diffuse gas \citep{tuellmann2009}.  

Most recently, a deep {\it Chandra} X-ray survey was carried out
covering the central 15$'$ of the galaxy, where the source density is
highest and {\it Chandra}’s exquisite spatial resolution is important
(T11). This survey used observations totaling 1.4 Ms to generate
a list of 662 sources.  Here we take advantage of the large field of
view and high soft sensitivity of {\it XMM-Newton} to produce a survey
complementary to T11.  Our survey is similar in depth, but covers the
full D$_{25}$ isophote and is more sensitive in the softest bands.  We
use this data set to identify new sources, look for pulsations in
known bright sources, and measure the radial source density for the
full extent of the galaxy. One source in the survey has been found to
be a transient X-ray pulsar \citep{trudolyubov2013}.

In this paper, we describe the methods we have used to extract a
catalog of source positions and fluxes from the data.  Producing a
reliable catalog turned out to be a significant technical challenge,
requiring adapting and customizing of many {\it XMM-Newton} science
analysis software (SAS) tasks.  As a resource to the community, we
therefore detail the procedures we have used to produce our catalog as
a significant fraction of this paper.  We have measured source
positions, position errors, detection likelihoods (DLs), count
rates, fluxes and errors in several energy bands.  From these results,
we obtained a catalog covering the entire M33 D$_{25}$ isophote and
beyond down to a limiting unabsorbed luminosity (0.2--4.5 keV) of
4$\times$10$^{34}$ erg~~s$^{-1}$, which we have used to measure
hardness ratios, search for short and long term variable sources,
classify soft sources, measure the radial distribution of sources, and
fit the XLF.

We have organized the paper as follows.  Section 2 describes the data
set; section 3 discusses the reduction of the data and source
detection technique in detail for others interested in analyzing
overlapping \XMM observations.  Section 4 details the resulting
catalog, including hardness ratios, comparisons with previous surveys,
and source variability. Section 5 discusses the properties of the
source population, including the radial source distribution and
XLF. Finally, Section 6 summarizes the conclusions.  All coordinates
in the paper are J2000.  We assume an inclination of 54$^{\circ}$
\citep{devaucouleurs1991} and a distance of 817 kpc
\citep{FreedmanDistance} for M33 throughout.  All reported fluxes are
unabsorbed.

\section{Observations}

\begin{figure}
\centerline{\epsfig{file=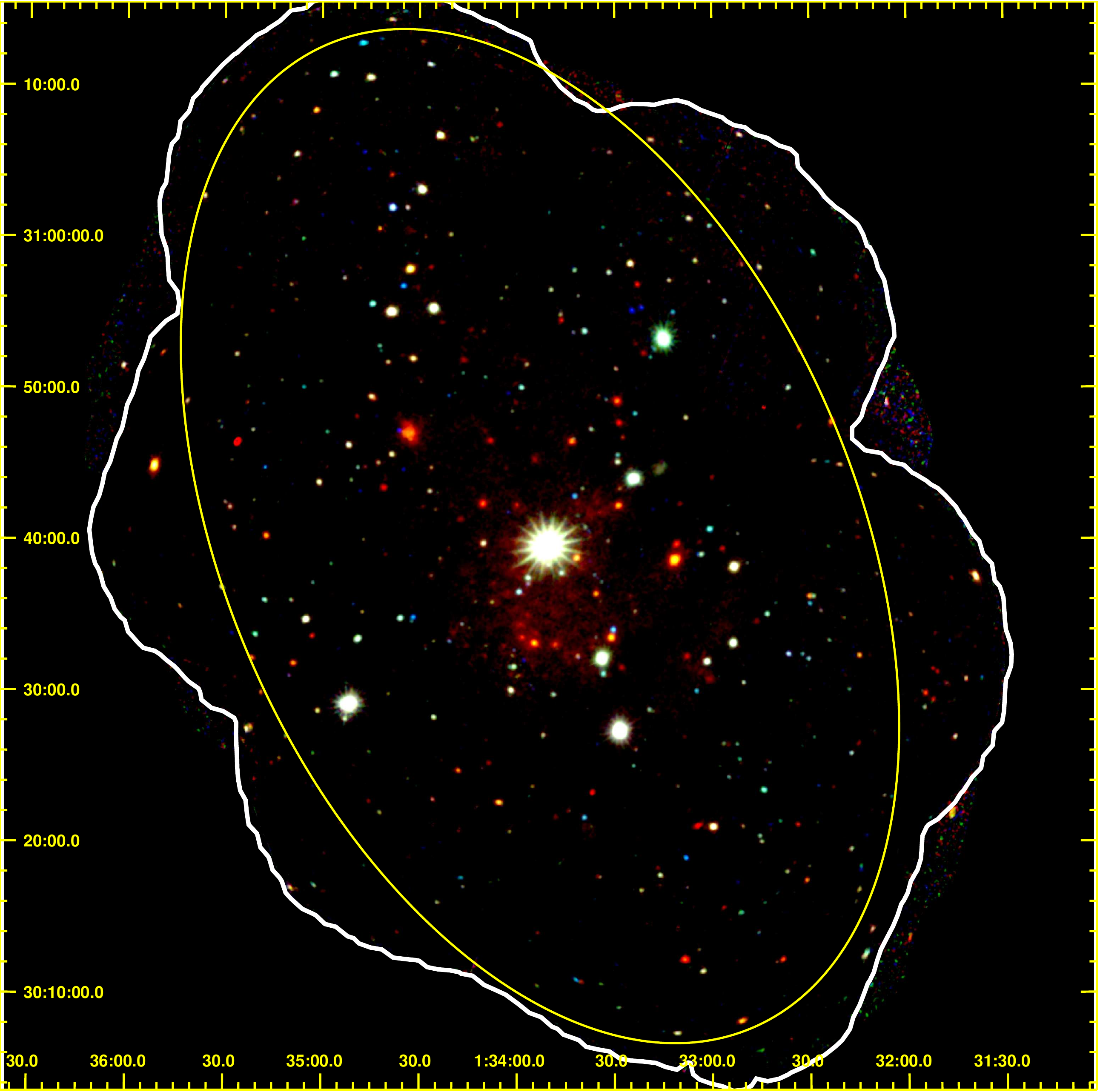,width=5.5in,angle=0}}
\caption{Our color composite image of M33 from all of the observations
listed in Table 1. Yellow ellipse marks the M33 D$_{25}$ region, and
the white contour marks the area with at least 40 ks of exposure.  Red
is 0.2-0.5 keV.  Green is 1-2 keV, and Blue is 2-4.5 keV.}
\label{image}
\end{figure}

\begin{figure}
\centerline{\epsfig{file=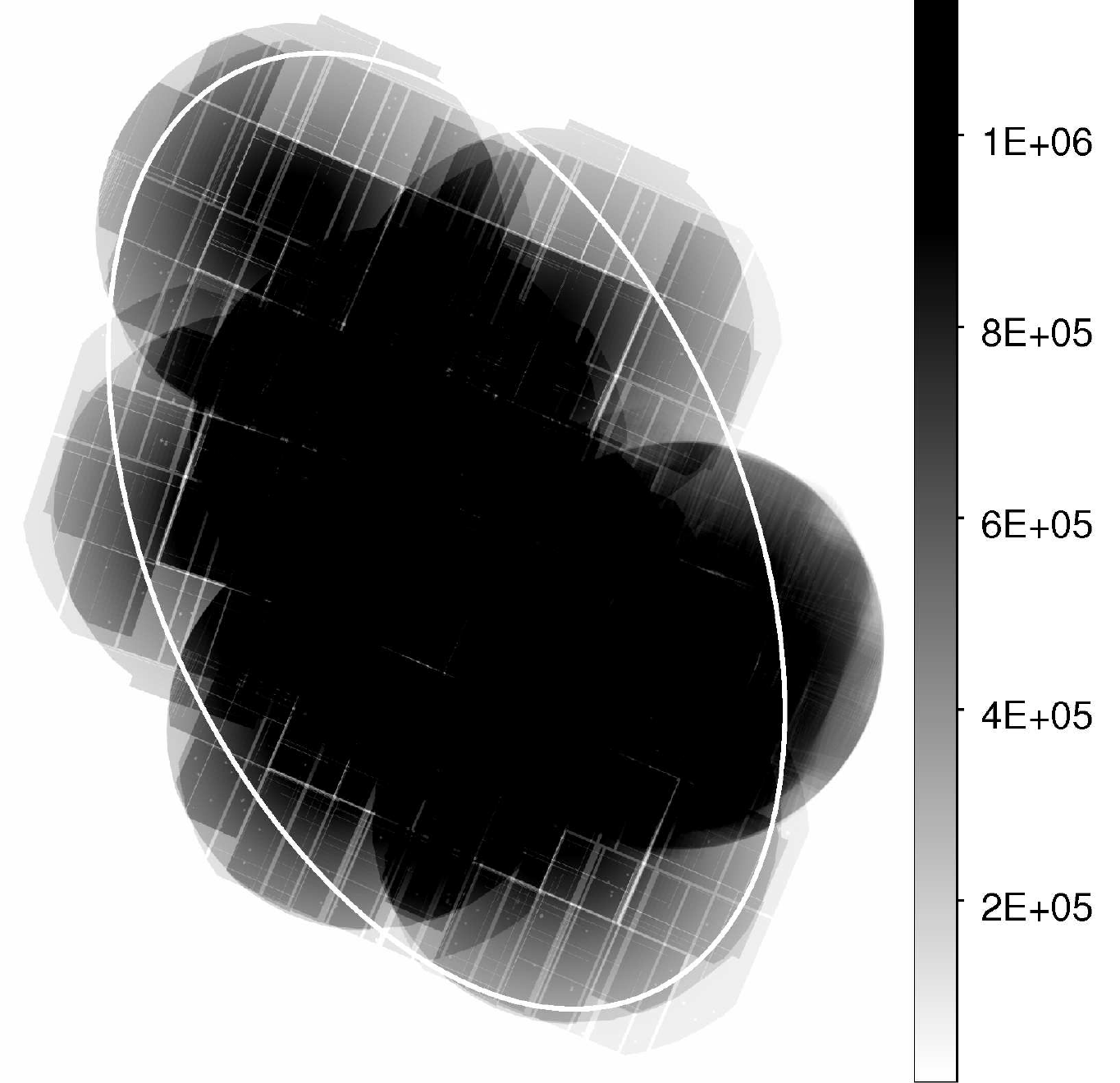,width=3.5in,angle=0}}
\caption{Our total exposure map (0.2-4.5 keV) from the survey data
  listed in Table 1.  This map was produced by {\tt emosaic\_prep} and
  includes the effects of vignetting.  The ellipse marks the M33
  D$_{25}$ region. The grayscale has units of summed PN, MOS1, and
  MOS2 seconds.}
\label{expmap}
\end{figure}

To produce an M33 catalog, we have used data of several newly observed
fields in M33 and archival observation of an eighth field.  The summed
exposures total about 900 ks (including background flare intervals).
Our new observations were designed to cover the whole D$_{25}$
isophote and provide overlap in conjunction with the archival data.
We show the color composite image and 0.2-4.5 keV exposure map in
Figures~\ref{image} and \ref{expmap}. The observation dates for the
seven new data fields ranged from 2010-07-09 to 2010-08-15 and from
2012-01-10 to 2012-01-12 (see Table~\ref{obstable}).  The observation
dates for archival data of the eighth field [PI: Pietsch], identified
by the prefix PMH (taken from the naming in P04 and M06) ranged from
2010-01-07 to 2010-02-24 to constrain the variability of the source
PMH~47.

{\it XMM-Newton} carries three cameras which together make up the
European Photon Imaging Camera (EPIC): two are comprised of metal
oxide semi-conductor \citep[MOS][]{turner2001} CCDs and the other is a
monolithic array of pn-CCDs \citep[PN][]{struder2001}. The PN CCDs
receive all incoming photons while the MOS instruments receive
$\sim$44\% of incoming photons due to their location behind a
grating which deflects some photons to
spectrometers\footnote{http://xmm.esac.esa.int/external/xmm\_user\_support/documentation/technical/EPIC/}. Most
{\it XMM-Newton} observations experience periods of background flaring
\citep{MOSbackground}; however, the original observation of field 4 was
affected by much higher flaring during the whole of the
observation. Therefore, we re-observed this field in January of 2012
so the entirety of the D$_{25}$ isophote is included in this study.

Our survey contains a total (after good time interval [GTI] filtering)
of 707.2, 707.1 and 680.5 ks of exposure for MOS1, MOS2 and PN
respectively.  The portions of these that come from the combined
PMH~47 observations after GTI filtering are 111.4, 111.4 and 99.8 for
MOS1, MOS2 and PN respectively.  Thus, we lost about 20\% of our
exposure time to background flaring.  In the next section, we will
detail how these data were analyzed to produce our source catalog.

\section{Data Reduction}

Our data reduction strategy was designed to maximize the reliability
and depth of the detected sources.  To accomplish this, we first
aligned all of the individual observations to the catalog of T11.  We
removed streaks from very bright sources in our fields. We filtered
out time intervals containing background flares using full-field
lightcurves.  Finally, we produced clean, astrometrically-aligned
images in each camera in four bands for pointing, along with matching
background maps and exposure maps for source detection and
characterization, as detailed below.

\subsection{Alignment}\label{alignment}

To correct the the boresight for each XMM observation, we used the
X-ray source catalog from the {\it Chandra} survey of M33 (T11) for
the reference system. The T11 catalog positions are aligned to within
0\farcs1 of the 2MASS and USNO-B1.0 all sky catalogs.  Each of our
long observations had from 23 to 79 matched sources to use for
alignment, while the short PMH 47 observations had from 4 to 27 such
sources.  For the source detection in this step we followed the method
described in \citet{haberl2012}. In a first run, we performed the
source detection using the SAS meta task {\tt edetect\_chain} for 15
images simultaneously from the three EPIC instruments in four
different energy bands 0.2-0.5 keV, 0.5-1.0 keV, 1.0-2.0 keV, 2.0-4.5
keV.  Background flares were removed by creating GTIs using the
technique described in Section~\ref{ps}, and out-of-time (OOT) events
for EPIC PN were taken into account as described in Section~\ref{oot}.
The resulting source lists from each observation were then correlated
with the {\it Chandra} catalog.  The inferred shifts in RA and Dec
were applied to the attitude observation data file (ODF) file for each
observation. The attitude file contains RA and Dec, so the shift is
simply added to the initial values.  The applied corrections are
typically between 0.5$''$ and 1$''$ in each coordinate, but in two
cases reached $\sim$4$''$ in RA. Such shifts are consistent with the
measured pointing accuracy of
\XMM.\footnote{http://xmm.esac.esa.int/external/xmm\_user\_support/documentation/uhb\_2.1/node108.html}. Once
the observations were aligned, all processing was rerun from the basic
ODF products using astrometrically aligned data products.  Finally,
once our full survey catalog was complete, we found we could improve
our absolute astrometric alignment with T11 by shifting the positions
of the entire catalog by $+$0.1$''$ in RA and $+$0.7$''$ in Dec, as
detailed in Section~\ref{matching}.

\subsection{Out-of-Time Events}\label{oot}

Out-of-time (OOT) events occurred in our observations because there
were bright sources in the field of view.  When such bright sources
are present, more than one photon can be registered for a pixel during
the readout of the CCD. Such events are thus given incorrect RAWY
values, creating stripes on the resulting image.  Furthermore, these
photons are given incorrect energy corrections for the charge transfer
inefficiency\footnote{http://xmm.esac.esa.int/external/xmm\_user\_support/documentation/sas\_usg/USG/},
which artificially broadens spectral features. 

To remove artifacts due to OOT events, we processed the raw data
(after alignment), in the form of ODFs, using SAS software. We
executed the SAS scripts {\tt emchain} and {\tt epchain} (for the
EPIC-MOS and EPIC-pn cameras, respectively), which ran a sequence of
SAS tasks that processed the ODFs into event lists and related files,
such as background lightcurves. We ran {\tt epchain} twice, first with
the parameter {\it withoutoftime=true} and then with it set to {\it
  withoutoftime=false}, creating the PN out-of-time (OOT) event list
without disrupting the original PN event list and related output
files.  By subtracting images made from the OOT list from images made
from the full event list, we cleaned the images, reducing the number
of spurious source detections due to OOT events.

\subsection{Background Filtering and Mapping}\label{gtis}

The \XMM mirrors have a high reflectivity efficiency for low energy
protons, resulting in a highly time-variable background.  Background
flares can produce count rates 1-2 orders of magnitude higher than the
quiescent level.  During such flares, data taken by {\it XMM-Newton}
are of little to no use.  Our goal was to make the source detection
GTIs relatively liberal, allowing us to find fainter sources without
being overly conservative about background contamination. 

\begin{figure}
\centerline{\epsfig{file=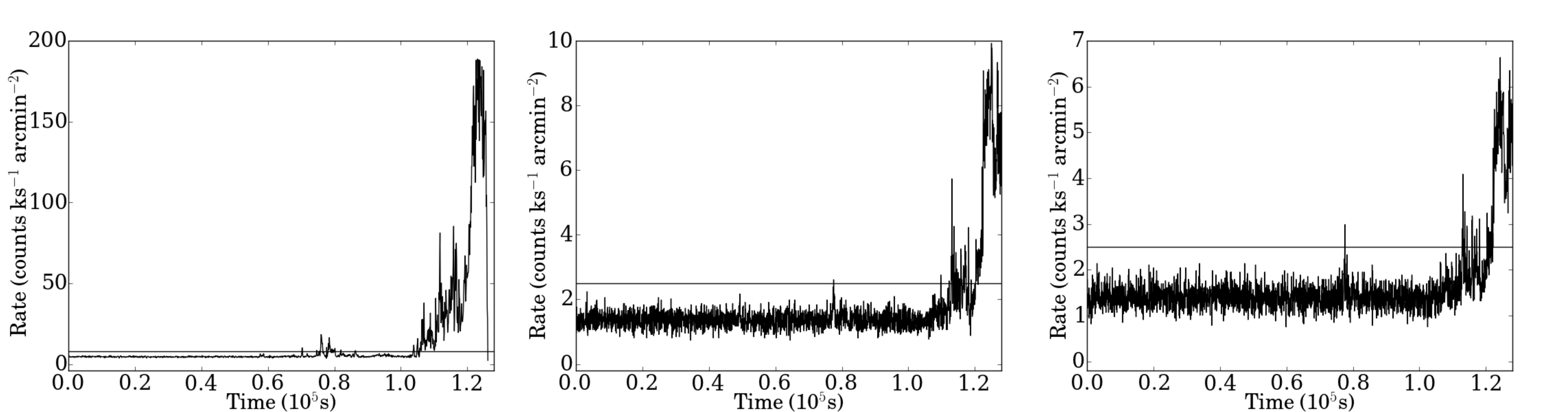,width=6.5in,angle=0}}
\caption{Light curves for Field 6, showing typical flaring.  Left is PN; middle is MOS1; right is MOS2.  Horizontal lines mark our cuts for determining GTIs.}
\label{rate_curves}
\end{figure}

We constructed GTIs from an examination of each observation's total
7--15 keV light curve in both MOS and PN, obtained by {\tt ep/emchain}
after removal of sources. Examples of these lightcurves are shown in
Figure~\ref{rate_curves}. General cutoff values were determined based
on the flaring levels in all observations, which led to a single
threshold value for each camera that we applied to all of the
data. For both MOS cameras, this value was 2.5 counts
ks$^{-1}$~arcmin$^{-2}$, and for the PN, the threshold was 8 counts
ks$^{-1}$~arcmin$^{-2}$ (both 7--15 keV).  The effective exposure
columns in Table~\ref{obstable} show the reduction in good time from
applying these GTIs.  

\begin{figure}
\centerline{\epsfig{file=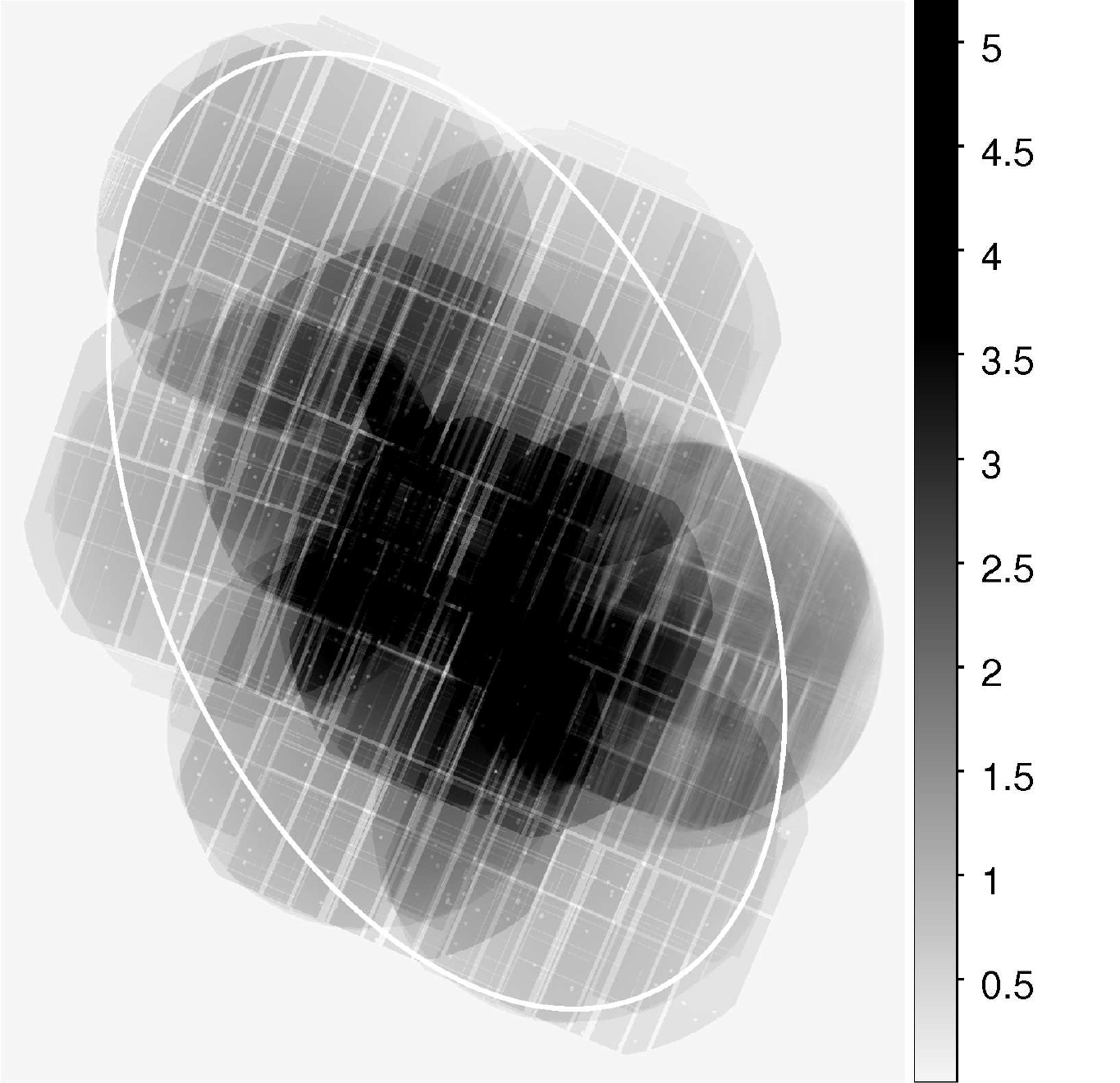,width=3.5in,angle=0}}
\caption{Our total background map (0.2-4.5 keV) from the survey data
  listed in Table 1.  The ellipse marks the M33 D$_{25}$
  region. Grayscale has units of counts pixel$^{-1}$ (2.25$''$
  pixel$^{-1}$).}
\label{background_map}
\end{figure}

We created background maps using these GTIs with the SAS task {\tt
  esplinemap}.  This task masks out all of the detected sources and
performs a spline fit over the remaining data to create a map of the
background.  For a survey-wide background map, we then summed the
resulting background maps of all observations
(Figure~\ref{background_map}).

\subsection{Event List Flagging}

In addition to GTI filtering, we flagged the data using FLAG==0 for
EPIC-MOS, which excludes all events at the edge of the CCD and
adjacent to bad pixels. For EPIC-pn, we applied (FLAG \& 0xfa0000)==0,
which provides a set of standard flagging options (e.g. events at edge
of CCD or outside the field of view).  We performed pattern filtering
with the selection criteria most appropriate for each instrument and
energy band as follows. EPIC-MOS filtering included single, double,
triple, and quadruple patterns. EPIC-pn filtering included single and
double patterns for energy bands 0.5-1.0, 1.0-2.0 and 2.0-4.5 keV, but
only single patterns for 0.2-0.5 keV. Patterns greater than double are
rare and have poor energy resolution. In addition, the PN detector
suffers from large numbers of spurious events with patterns greater
than single in the softest band making more conservative filtering
necessary.

\subsection{Source Detection}\label{ps}

To prepare our background maps and filtered event lists for
  source detection, we applied the SAS scripts {\tt emosaic\_prep} and
  {\tt emosaicproc}, which performed initial source detection on our
full stack of observations simultaneously.  The script was not
specifically designed to work with multiple partially-overlapping
observations.  We therefore had to make some adjustments to both our
data and the script.  For example, some of our data sets' keywords
were the same for different observations, which produced an
error. Thus, we explicitly confirmed such keywords of the output files
of {\tt emosaic\_prep} were unique.  With the images and exposure maps
from {\tt emosaic\_prep}, our files were set up to run {\tt
  emosaicproc}, which included the event lists from all three
instruments from all observations.

Next, we ran {\tt emosaicproc} on all of the prepared images and maps,
which applies the task {\tt eboxdetect} to produce a list of candidate
source positions.  We performed {\tt eboxdetect} runs using the
default parameters, other than the following changes to optimize it
for our data set. We set usemap=true, which uses map mode instead of
local mode; set withdetmask=true, which uses the detector masks we
produced in {\tt emosaic\_prep}; set withexpimage=true, which uses the
exposure maps we produced with {\tt emosaic\_prep}; set nruns=1, as
recommended in {\tt emosaicproc}; and finally, set likemin=4 to be
more inclusive in the positions provided to {\tt emldetect}.

Once we had determined all of the candidate source positions with {\tt
  emosaicproc}, we ran the SAS task {\tt emldetect} on the output list
from {\tt eboxdetect} with no cuts to the candidate catalog, allowing
source splitting and repositioning in order to take advantage of the
ability of {\tt emldetect} to fit the {\it XMM-Newton} point spread
function (PSF).  Then, we iterated running {\tt emldetect} on the
resulting output source list, but keeping the source positions fixed
and not allowing further source splitting, until the results of {\tt
  emldetect} converged on a repeatable solution.

We next determined appropriate energy conversion factor (ECF)
values. {\tt Emldetect} converts the measured off-axis count rates
into equivalent on-axis count rates using a model of the detector
response and vignetting function. Thus, the count rates reported are
on-axis equivalent.  These on-axis rates are converted to the reported
fluxes using the ECFs appropriate for the spectral model we selected.
We determined our ECF values for the different bands and instruments
using XSPEC \citep{arnaud1996}, assuming a power-law spectrum with
photon index 1.7 and absorption 6$\times$10$^{20}$ cm$^{-2}$, matching
M06.  We assumed this spectrum for all count rate to flux conversions
in this project.  We extracted one on-axis ARF for each band-camera
and applied it to our model spectrum in XSPEC to determine our ECFs.
The ECFs provide the unabsorbed fluxes assuming this spectrum and are
listed in Table~\ref{ecfs}.

\begin{figure}
\centerline{\epsfig{file=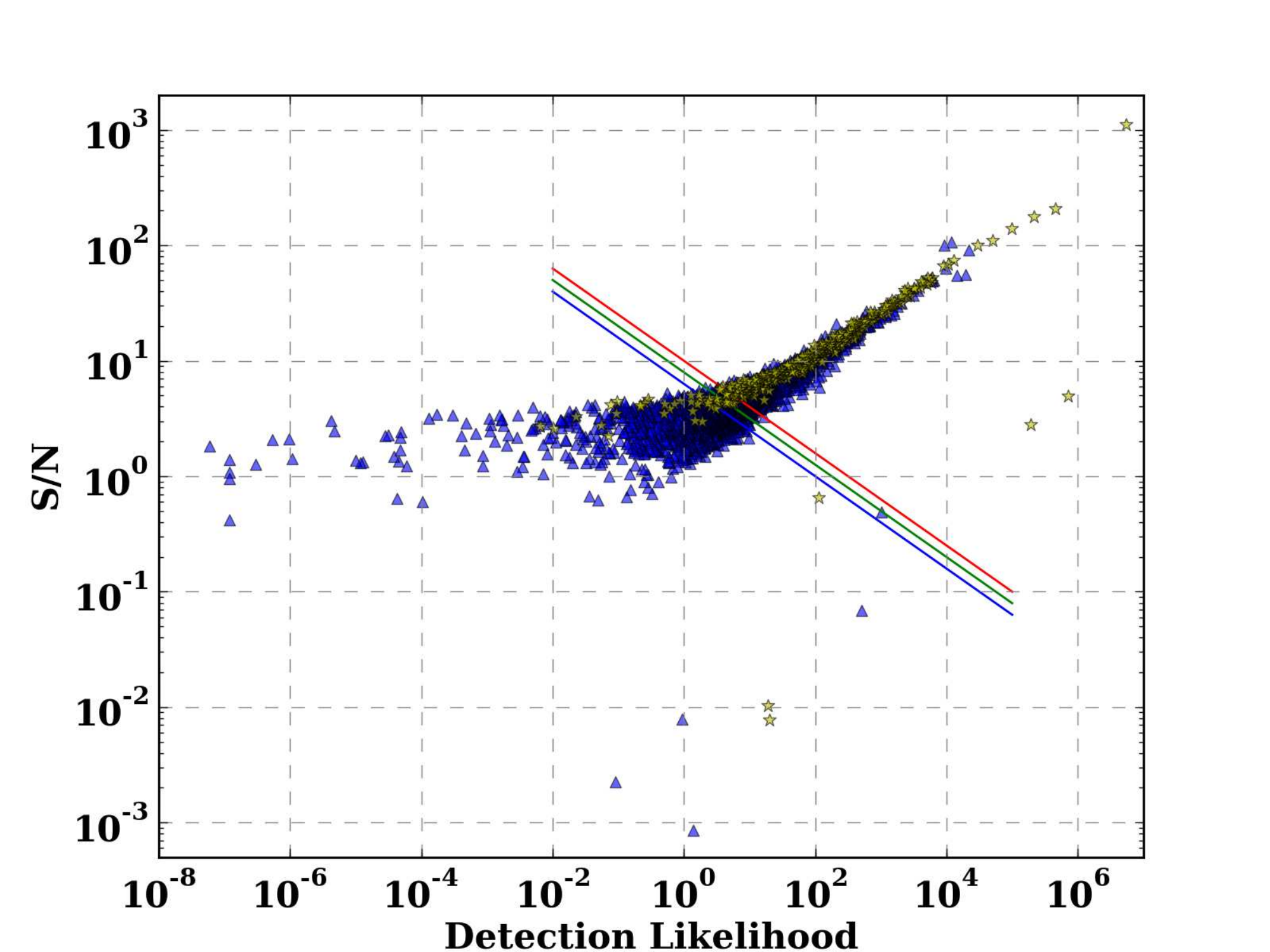,width=3.0in,angle=0}}
\centerline{\epsfig{file=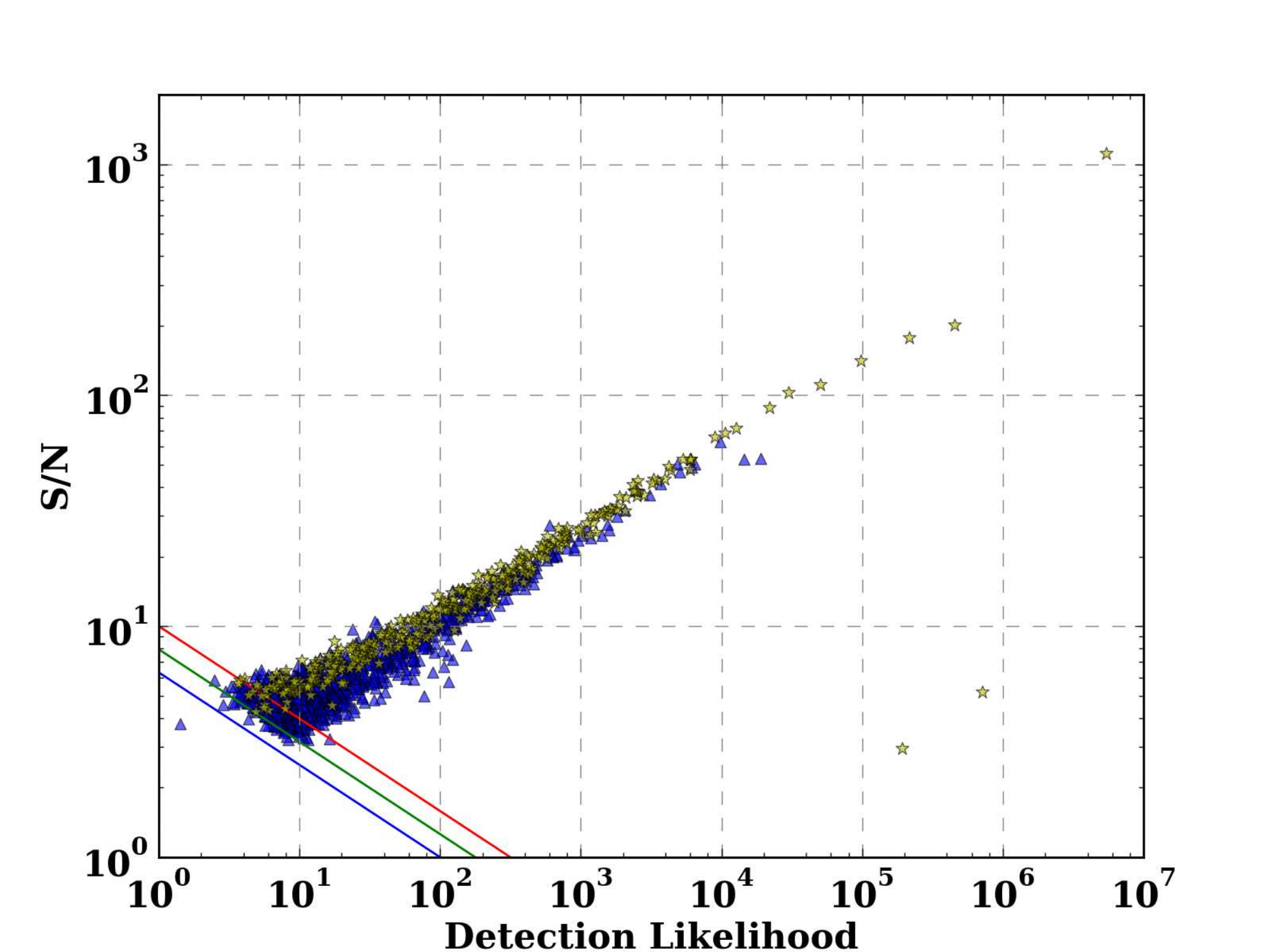,width=3.0in,angle=0}}
\centerline{\epsfig{file=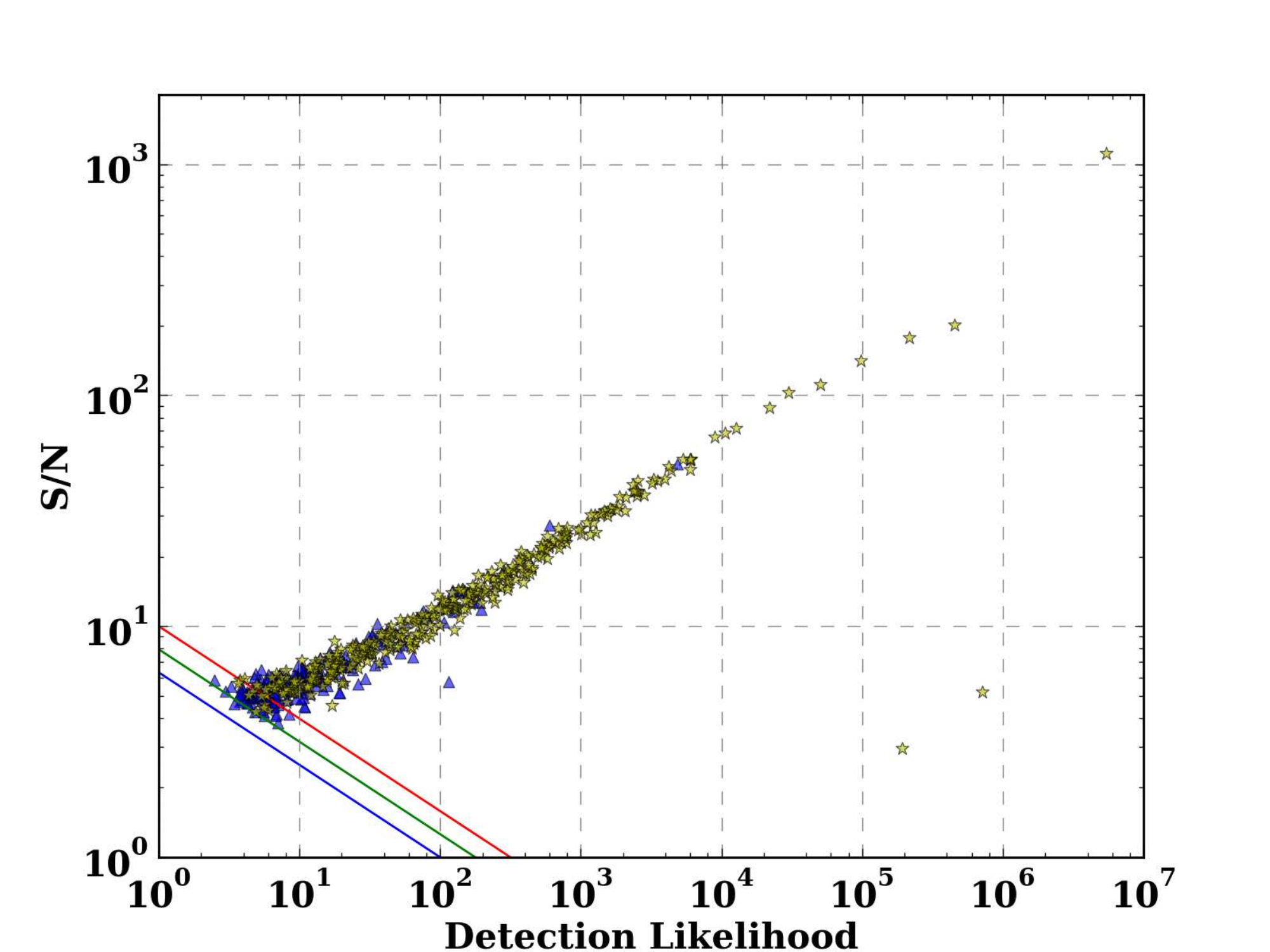,width=3.0in,angle=0}}
\caption{{\it Top:} Signal-to-noise vs. DL for all of
  the source candidates from {\tt eboxdetect} prior to making our
  quality cut.  Blue triangles mark the measurements when data from
  all instruments from 0.2-4.5 keV are included.  Stars mark sources
  matched with the catalog of T11, and diagonal lines
  mark possible quality cuts.  The green line marks the cut we used
  for the final catalog. {\it Middle:} Same as {\it top} for our
  final catalog. {\it Bottom:} Same as {\it top}, but for our final
  catalog and showing only sources inside the T11
  footprint. }
\label{cuts}
\end{figure}

With the full list of candidate positions reliably characterized by
{\tt emldetect}, we determined a quality cut to remove spurious
detections using the signal-to-noise (counts/counts\_error) and
DL values from the total
(0.2-4.5 keV) band.  A plot showing our method is provided in
Figure~\ref{cuts}.  We located a subsample of previously known sources
by eye using the T11 catalog.  We plotted these sources with
yellow stars on a plot of all of the signal-to-noise and DL values.
The two values are clearly correlated, as seen in Figure~\ref{cuts};
however, the standard DL$>$6, corresponding to a null
probability of 0.0025,\footnote{The DL is calculated as $-ln(p)$
  where $p$ is the probability that no source is present, where
    $p=1-P(\nu/2,L')$, with $L'=\sum\limits_{i=1}^n C_i/2$ ($C$
    defined by \citealp{cash1979}), ${\nu}=2+n$, and $P$ is the
    incomplete Gamma function. Here, $n=1$ for an individual band in
    an individual
    observation. (http://xmm.esac.esa.int/sas/current/doc/emldetect.pdf)}
would remove many previously known sources from our catalog.  These
sources have DL$<$6, but relatively high signal-to-noise ratios
($\gap$4).  These are likely the result of the M33 diffuse
  background and crowding of sources making background fitting and
  PSF fitting less accurate.  Therefore we chose
our cut using a line perpendicular to the signal-to-noise vs. DL
correlation such that we included portions with a large fraction of
known sources, but excluded low-confidence sources.  Our final cut
corresponded to log(S/N)${\leq}-0.4$log$({\rm DL})+0.9$.

We verified the reasonableness of this cut by checking that it did not
exclude any sources that were easily seen by eye.  We note the few
outlying sources with high DL but low S/N values in the top panel of
Figure~\ref{cuts}.  There are five of these with 10$<$DL$<$10$^4$.
Three were single sources that each had multiple entries in the
preliminary catalogs, which resulted in one of the entries being given
a very low S/N.  We forced each of these three to each be measured as
single sources.  Their final measurements (Sources 603, 773, 864) lie
on the relation.

Two other outliers were very bright sources (484 and 521), and
continue to have large count uncertainties leading to the low S/N
values in the bottom panel of Figure~\ref{cuts}.  These two sources
have large count uncertainties for different reasons.  Source 484 lies
in the center of a PN chip gap in one observation, so that only the
wings of the PSF could be fitted, likely affecting the error
computation.  Source 521 is M33 X-7, the eclipsing binary, which had
very few counts in one observation where it was in eclipse, likely
affecting the error computation.  Their flux uncertainties as reported
by {\tt emldetect} are low, as would be expected for such bright
sources.  The weighting used for computing counting errors and flux
errors are different. For flux uncertainties, the most weight is given
to the band-camera images with the most counts, but for counting
uncertainties, 
  the count-rate errors are weighted by exposure and combined.  The difference in
  weighting causes the difference in fractional error between counts
  and flux for these two sources.  We also note that no weighting
    is performed in the combination of DL values from multiple bands.
    All the parameters that determine the available counts from a
    source (and background) go into the likelihood value so that the
    likelihood values can be simply added together (properly
    normalized to the same number of degrees of
    freedom).\footnote{http://xmm.esac.esa.int/sas/current/doc/emldetect.pdf}

After applying our quality cuts, we ran {\tt emldetect} on the
culled catalog, allowing the positions to be refitted, without being
influenced by the 1703 likely artifacts that were culled.  Thus, our
final catalog positions have been determined after the culling of
positions unlikely to contain a real source. {\tt Emldetect} has a
parameter, {\it fitnegative}, which defaults to no, thereby invoking a
constraint that excludes PSF fits with negative fluxes.  This
constraint forces the minimum allowed measured flux to be zero and
forces counts only to be subtracted from (never added to) the data for
each fitted source during the measuring process. For such
  non-detections, {\tt emldetect} reports the 1$\sigma$ upper limit
  for the source counts, count rate, and flux, which corresponds to
  the 68th percentile of allowed positive PSF fits in {\tt emldetect}. Thus all
  uncertainties for values of zero in our source catalog should be
  treated as 1$\sigma$ upper limits. This is the standard
implementation of the fitting, as used by the 3XMM catalog
\citep{3xmm2013}. We chose to keep the default option for our catalog.
However, there are 40 sources in our catalog that have zero flux in
any band, so that apparently only a small fraction of the catalog
fluxes were forced to zero because of this fitting technique.  For
these sources, the uncertainty gives the 1$\sigma$ upper limit as
  reported by emldetect.

Once we had our final list of source positions, we ran {\tt emldetect}
simultaneously on the 0.2-0.5, 0.5-1.0, 1.0-2.0, and 2.0-4.5 keV bands
including all data in order to determine the total signal-to-noise and
DL.  These values come from the total combination line for each
source that is output from {\tt emldetect} when all of the data were
included in a single run.  Thus, we did not specifically run a 0.2-4.5
keV band, but instead used the combined total from the four individual
bands for our ``total'' (0.2-4.5 keV) measurements.  The
  combination of the measurements in different bands is performed
  within emldetect, weighting each pixel in each image appropriately
  based on the exposure given the \XMM calibration files, as well as
  the background images and masks for each observation.  These weights
  are not based on count rates.  The counts are added and divided by
  the total exposure.  Thus, if a source is much fainter in one
  observation than another, the weighting of the lower-count-rate
  observation will not be lower, and could be higher if it was an
  observation with a longer effective exposure.

In addition, we ran the final list of positions through {\tt
  emldetect} including all observations four more times to measure the
characteristics of the sources in each individual band.  The fluxes
and count rates from these runs were used for measuring hardness
ratios.  Finally, we ran {\tt emldetect} on the same list in each band
{\it including only one camera at a time}.  These runs provided fluxes
and count rates in each camera in each band for the catalog.

Furthermore, we searched for false detections attributable to hot
pixels or OOT streaks.  Our technique was to look at the sources by
eye in each individual observation, on soft-band images produced in
detector coordinates.  The soft band was chosen because it is the most
susceptible to hot pixel artifacts \citep{haberl2012}.  We flagged
(see Column 9 description in Section~\ref{cattext}) all sources that
corresponded to columns aligned with bright sources and sources that
appeared coincident with hot pixels in the soft band images.  We also
looked at any catalog entry with a PN 0.2-0.5 keV DL value that was
much higher than the value in any other band/camera (DL$_{\rm
  PNsoft}{>}5$, with all other band cameras DL$<$DL$_{\rm PNsoft}/2$).
Any of these entries that did not appear to be a convincing source in
the combined image was flagged as a hot PN pixel in the final catalog
(column 9).

In addition, the MOS CCDs are known to suffer from occasional high
noise levels (see {\it XMM-Newton} Calibration Technical Notes,
released periodically through the
ESA\footnote{http://xmm2.esac.esa.int/docs/documents/CAL-TN-0018.pdf}),
which can lead to spurious source detections.  To remove such
artifacts, we looked at all entries that had a higher DL value
in any MOS1 or MOS2 band than in all other camera/bands (for example,
DL$_{\rm MOS1band}{>}5$, DL$_{\rm PNband}{<}$DL$_{\rm
  MOS1band}/5$, DL$_{\rm MOS2band}{<}$DL$_{\rm MOS1band}/5$, and
all other band cameras DL$<$DL$_{\rm MOS1band}$).  Any of these
locations that did not appear to be a convincing source in the
combined image was flagged as a MOS artifact in the final catalog.  In
total, we flagged 75 sources as soft-band PN or OOT artifacts and 138
sources as MOS artifacts, with 4 sources flagged as both.  Fifteen of
these flagged measurements were matched to T11 sources and seven to
M06 sources (with three matched to both), which suggests they are more
likely to be useful measurements. In all, we flagged 209 sources as
possible artifacts, with 190 of these being unmatched to any previous
survey.  Our analysis for this paper includes flagged sources that had
T11 matches, but excludes all other flagged sources.

We also found that when sources were not detected in all cameras in a
band, then {\tt emldetect} did not compute a combined flux for the
band.  In these cases, we flagged the source (column 9 in the
catalog), and computed our own combined flux from the cameras where
the source was detected (see Section~\ref{cattext}).  Furthermore,
about 3\% of the sources had combined measurements across all bands
(0.2-4.5 keV) in {\tt emldetect} that differed significantly from the
summation of independent measurements made in each band independently.
While the veracity of these sources is not an issue, the measurement
discrepancies suggest that their fluxes are less reliable than the
other 97\% of the catalog, and therefore we gave them flags described
in the next section.

Finally, we ran the SAS task {\tt esensmap} using our exposure and background
maps to produce sensitivity maps.  These sensitivity maps were needed
for our XLF analysis, described in Section 4.3.

\section{Results}

\subsection{Source Catalog}\label{cattext}

Our source catalog with a total of 1296 sources
is provided in Table~\ref{catalog}.  The columns are:

\begin{itemize}

\item{Column 1: Source Identification Number}

\item{Column 2-3: Right Ascension and Declination in J2000 coordinates}

\item{Column 4: Position Uncertainty in arcseconds}

\item{Column 5: DL as $-ln(p)$, where $p$ is the probability that no source is present, including all of the observations (all bands, all cameras) measured.}

\item{Column 6: The net source counts and uncertainty combined across
  all cameras and bands (0.2-4.5 keV) measured by simultaneously
  fitting the EPIC PSF to all observations of the source in all bands
  and cameras.}

\item{Column 7: The on-axis count rate and error combined for all bands (0.2-4.5 keV) and all cameras.}

\item{Column 8: The unabsorbed energy flux combined for all bands (0.2-4.5 keV)
  combining all measurements in our separate bands and cameras with their
  respective energy conversion factors based on a single on-axis
    ancillary response function (ARF) in each band-camera assuming a
    power-law spectrum of index 1.7 and an N$_{\rm
      H}$=6$\times$10$^{20}$ cm$^{-2}$ (see Table~\ref{ecfs}).}

\item{Column 9: Flag indicating any deficiencies with the detection or
  characterization.  Values are: ``0'' for no problems (904 sources),
  ``ip'' for a few hot pixels in an image (27 sources), ``is'' for an
  associated streak from a nearby bright source (29 sources), ``m''
  for high DL in one MOS camera but not the other MOS camera or
  the PN (138 sources), ``pn'' for much higher DL in the 0.2-0.5
  keV PN data than in any other band-camera (35 sources). Such
    sources are likely to be spurious unless they have a match in
    another survey.  The ``s'' flag denotes a source that was not
    measured in all 3 cameras (167 sources).  In these cases, the
    total measurements are repeats of the measurements from the camera
    that detected the source (for single camera detections) or
    weighted mean measurements from the two cameras that detected the
    source.  These sources are not likely spurious, but their combined
    measurements were not performed by {\tt emldetect}.  Finally
  ``t'' flags indicate sources with 0.2-4.5 keV source counts that are
  more than a factor of 2 different from the sum of the source counts
  measured in the individual band-cameras (suggesting a problem with
  the merged measurement), or with total counts in an individual band
  more than a factor of 2 different from the sum of the source counts
  measured in the individual cameras for that band, likely due to a
  limitation of the combining algorithm.  Thus, for the 44 t-flag
  cases, the sources themselves are not spurious, but the total
  combined measurement are not as reliable.  Thus, we do not use the
  combined totals for analysis here.  We only analyze the individual
  band (or band-camera) measurements, as {\tt emldetect} produced
  inconsistent results in the stack of all band-camera data at these
  locations.}

\item{Column 10: Matching T11 source name. (see Section~\ref{matching})}

\item{Column 11: Matching M06 source name.  (see Section~\ref{matching})}

\item{Column 12: Secondary matched source, if a second T11 source was
matched (indicating a blend in our data).}

\item{Column 13: The source type if known.  This column indicates sources that are known supernova remnants or foreground stars based on previous studies or our own comparisons with optical data.}

\item{Columns 14-17: Same as columns 5-8, but for the 0.2-0.5 keV band
alone, combining the data from all observations from all cameras.}

\item{Columns 18-21: Same as columns 14-17, but for the 0.5-1.0 keV band.}

\item{Columns 22-25: Same as columns 14-17, but for the 1.0-2.0 keV band.}

\item{Columns 26-29: Same as columns 14-17, but for the 2.0-4.5 keV band.}

\item{Column 30: Total exposure time at the source location in the PN camera
in the 0.2-0.5 keV band. Exposure time is computed in each band
separately, as vignetting is energy dependent.}

\item{Columns 31-34: Same as columns 14-17, but for all observations from
the PN camera alone.}

\item{Column 35-39: Same as columns 30-34, but for the 0.5-1.0 keV band.}

\item{Column 40-44: Same as columns 30-34, but for the 1.0-2.0 keV band.}

\item{Column 45-49: Same as columns 30-34, but for the 2.0-4.5 keV band.}

\item{Columns 50-69: Same as columns 30-49, but for all observations from MOS1.}

\item{Columns 70-89: Same as columns 30-49, but for all observations from MOS2.}

\item{Columns 90-91: Hardness ratios from fluxes; HR1 and HR2, as
  described in Section~\ref{hrs_text}.}

\item{Columns 92-93: Hardness ratios from source counts in the softest
  bands; HR$_{1C}$ and HR$_{2C}$ as described in
  Section~\ref{hrs_text}.}

\end{itemize}

\subsection{Hardness Ratios}\label{hrs_text}

Hardness ratios (HRs) compare fluxes across different X-ray bands
  and provide additional information about the types of sources
  contained in the catalog. To investigate the HRs present in our
  catalog, we computed two HRs using the source fluxes from our four
energy bands for all unflagged sources, sources with the ``s'' or
``t'' flag, and flagged sources matched to T11. All fluxes are
unabsorbed and assume a power-law spectrum with index 1.7 and
absorption 6$\times$10$^{20}$ cm$^{-2}$.  For sources with the ``s''
or ``t'' flag, we adopted the PN fluxes for each band. If the
  measured rate or flux is 0.0 in any of the bands, we adopt the upper
  limit from {\tt emldetect} (see Section~\ref{ps}) as the rate or
  flux for that band when computing the hardness ratios. The 0.2-0.5
  and 0.5-1.0 keV bands were combined into one soft (S) band, while
  1.0-2.0 and 2.0-4.5 keV made up the medium (M) and hard (H) bands
  respectively.  These bands make our HRs easier to compare with other
  surveys. The formulas we used for our HRs were

\[HR1 = \frac{M-S}{S+M+H},  HR2 = \frac{H-M}{S+M+H}\]

\noindent
where H, M, and S represent the fluxes of the sources in the three
bands defined above.  
 
\begin{figure}
\epsfig{file=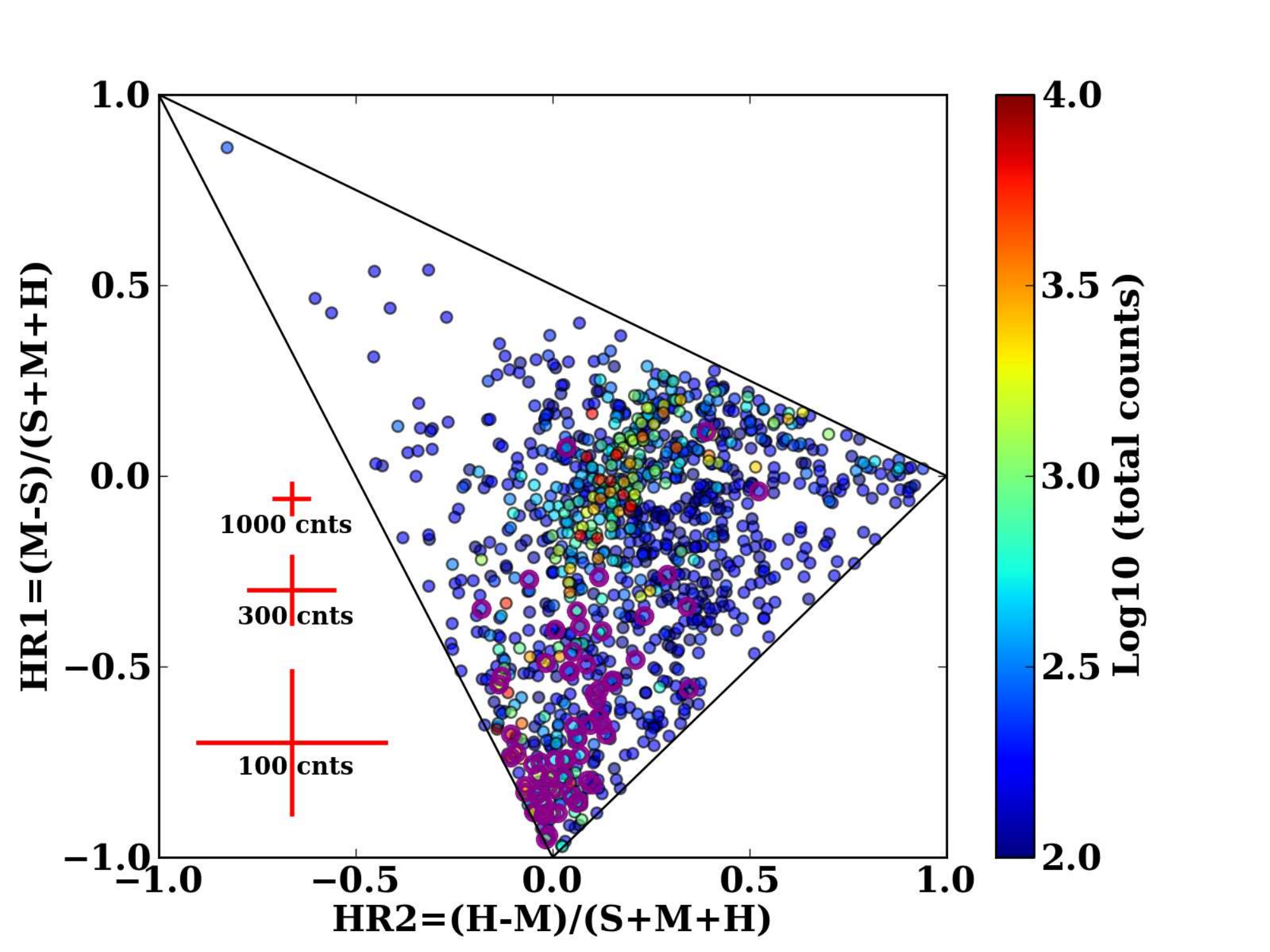,width=3.2in,angle=0}
{\hbox{\hspace{-4.7cm}\centerline{\epsfig{file=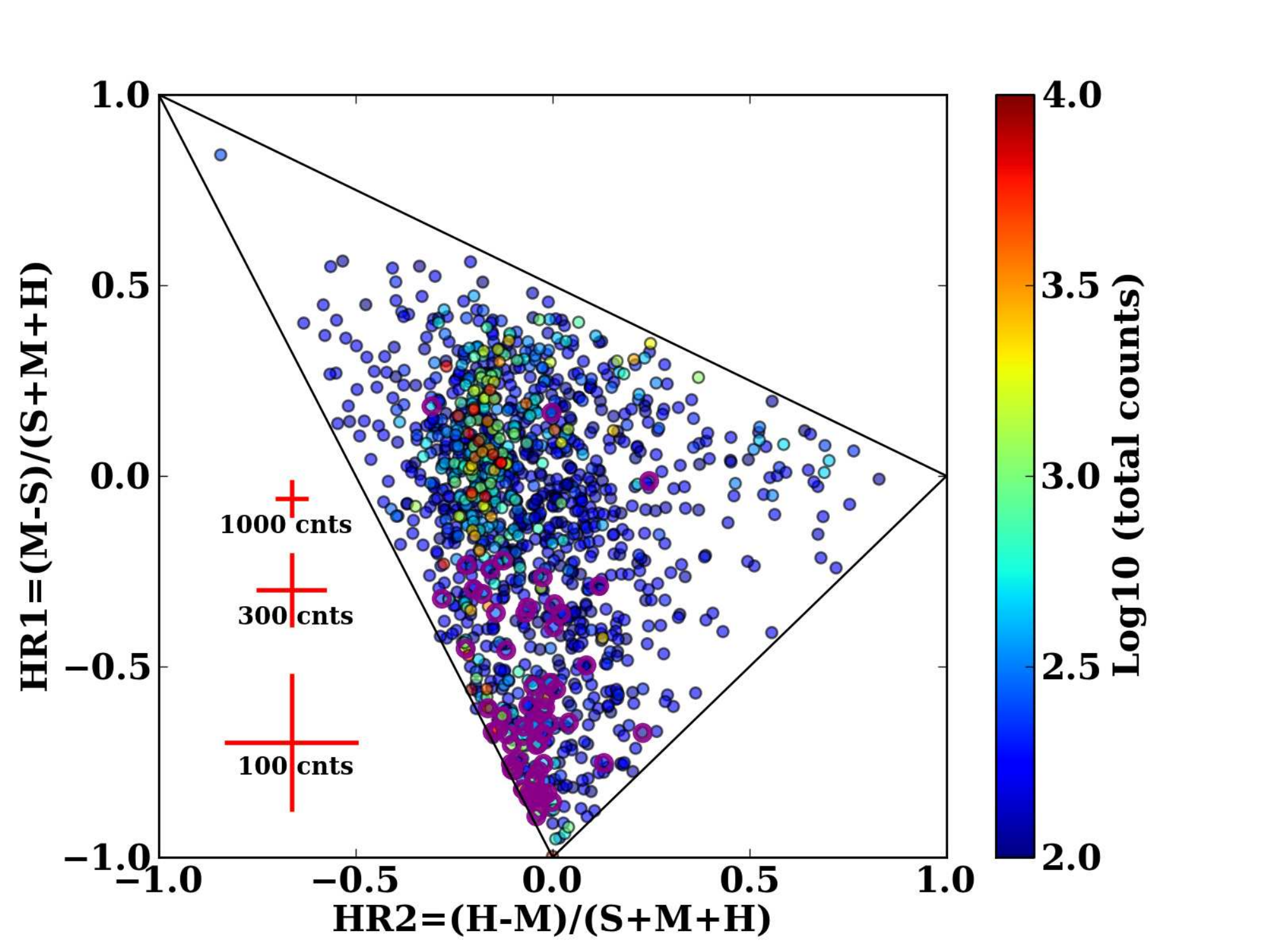,width=3.2in,angle=0}}}}
\epsfig{file=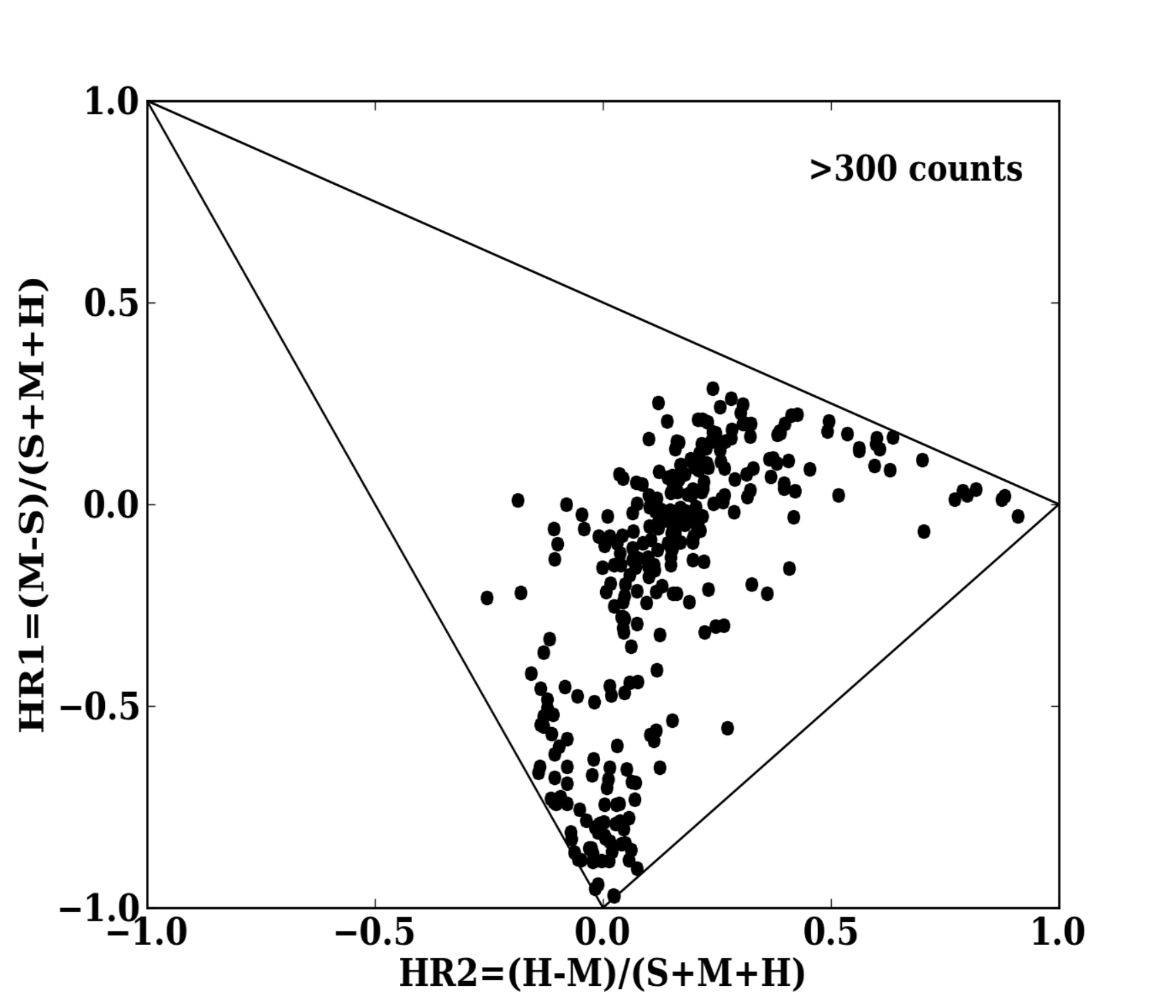,width=3.2in,angle=0}
\epsfig{file=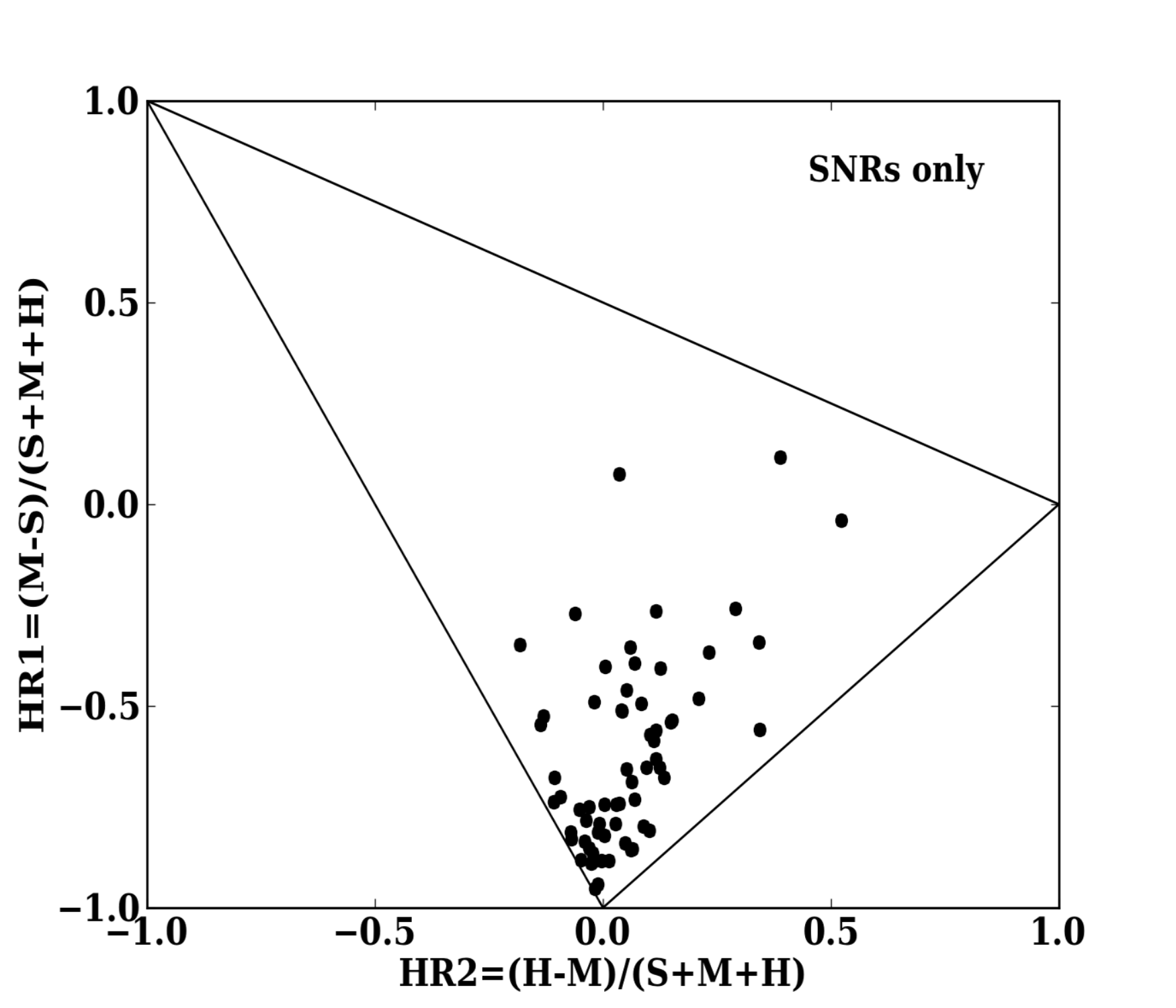,width=3.2in,angle=0}
\caption{{\it Top left:} Hardness ratios from fluxes in the
    0.2-1.0, 1.0-2.0, and 2.0-4.5 keV bands of all unflagged,
    s-flagged, t-flagged, or matched to T11 sources with $>$50 source
    counts in Table~\ref{catalog}.  These fluxes use the ECFs in
    Table~\ref{ecfs}, which assume a power-law spectrum with index 1.7
    and absorption 6$\times$10$^{20}$ cm$^{-2}$.  The 0.2-1.0 keV
    fluxes are the sum of the 0.2-0.5 keV and 0.5-1.0 keV fluxes in
    Table~\ref{catalog}. Fluxes are used for the HRs in this plot to
  make our catalog easier to compare with those measured with other
  X-ray observatories.  H = 2.0-4.5 keV flux; M = 1.0-2.0 keV flux;
  S=0.2-1.0 keV flux.  Upper limits were applied in bands with 0
  counts to calculate ratios, so that no sources fall outside of the
  area allowed by positive flux measurements.  Blue to red colors show
  the number of counts in the measurement.  Known SNRs are highlighted
  with purple circles. Approximate uncertainties for a range of total
  counts are shown with the red crosses.  {\it Top right:} Same
    as top-left, but ratios were calculated from count rates. {\it
    Bottom left:} Same as top-left, but without color-coding and
  showing only sources with $>$300 source counts in our catalog are
  shown.  The sequence from the soft thermal sources to the hard,
  absorbed power-law sources is much clearer. {\it Bottom right:} Same
  as top-left, but without color-coding and showing only known SNRs,
  which have a markedly softer distribution than the full source
  population.}
\label{HRs}
\end{figure}

We plot the resulting HRs in Figure~\ref{HRs}.  We include only
  sources with $>$50 source counts.  Because upper limits are used for
  non-detections, no sources fall outside of the area allowed by
  positive flux measurements. Fluxes are used for the HRs in three of
  the plots in Figure~\ref{HRs} to make our catalog easier to compare
  with those measured from other X-ray observatories. The upper right
  shows the plot as calculated from straight count rates.  Because the
  bands were measured independently, the fluxes are simply the
  count-rates multiplied by the ECFs in Table~\ref{ecfs}.  Thus,
  relative offsets between source positions are similar in both of the
  top panels.

 In Figure~\ref{HRs}, SNRs are highlighted with purple circles.
 Points are color-coded by number of counts to give a sense of the
 precision of the measurement.  The HRs congregate around a sequence
 from HR1${\sim}{-}$0.3 to HR1${\sim}{+}$0.3.  The SNRs stand out as
 soft (HR1${<}{-}$0.3).  In the bottom-left panel, we show only
   sources with $>$300 source counts in our catalog to highlight the
   sequence from the soft thermal sources to the hard, absorbed
   power-law sources.  In the bottom-right panel, we show only known
   SNRs, which have a markedly softer distribution than the full
   source population.

\begin{figure}
\centerline{\epsfig{file=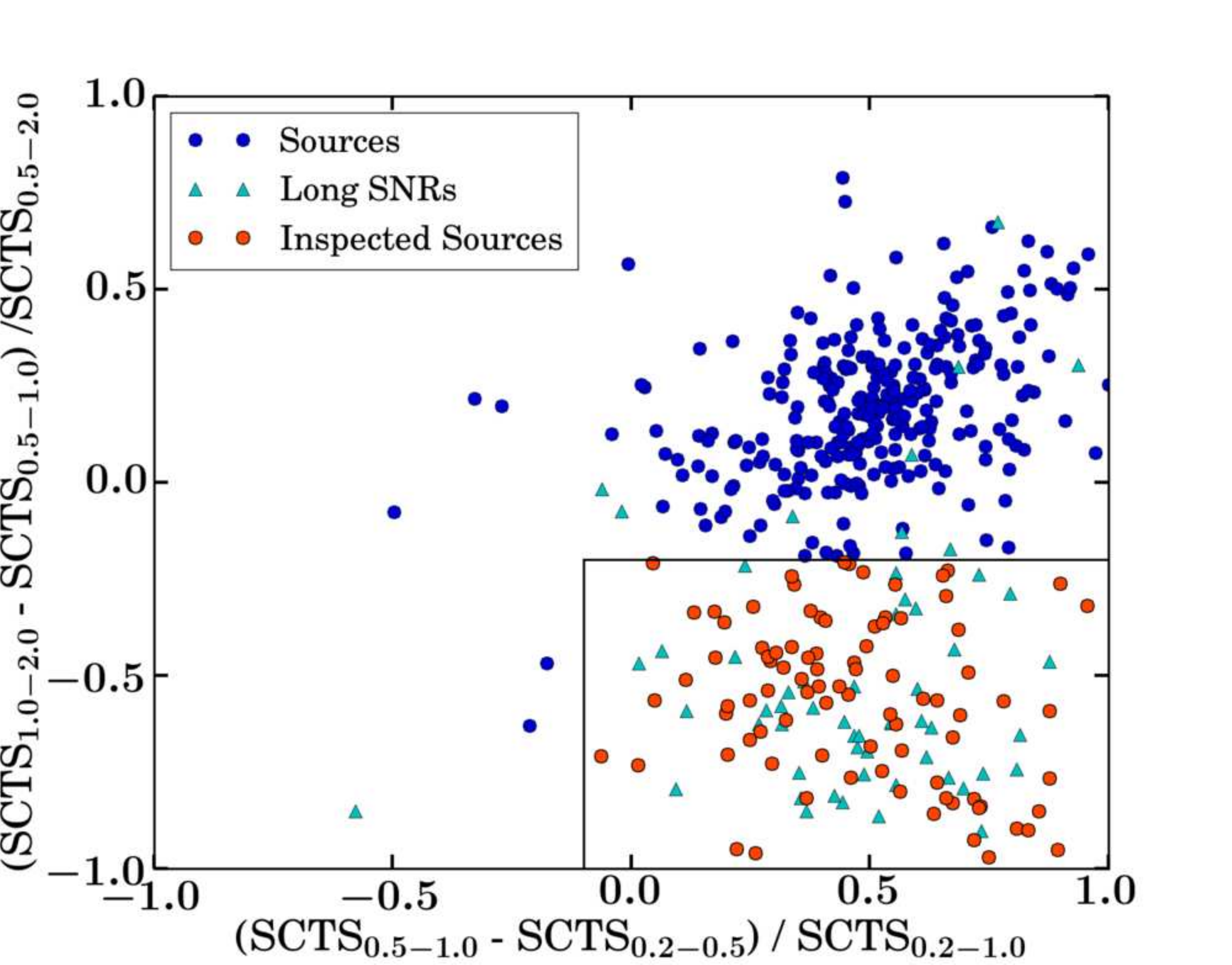,width=4.5in,angle=0}}
\caption{Hardness ratio from source counts (SCTS) in the 0.2-0.5,
  0.5-1.0, and 1.0-2.0 keV bands of all unflagged, s-flagged,
  t-flagged, or matched to T11 sources in our catalog with hardness
  ratio uncertainties $\leq$0.2 based on the number of counts in these
  bands.  We use these ratios with the counts to make the same
  comparisons as \citet{pietsch2004} for separating SNRs and
  foreground stars from sources with harder spectra.  The black box
  indicates the region where most \citet{LongSNR} SNRs (triangles)
  lie.  The other sources in this box, indicated by the 89 orange
  dots, were not in the \citet{LongSNR} catalog.  These 89 were all
  investigated in optical images, and classified as SNRs if they
  coincided with gas emission or foreground stars if they coincided
  with a bright star.  Sixty-two of the 89 remain unclassified.}
\label{KristenHR}
\end{figure}

In Figure~\ref{KristenHR}, we show the hardness ratios based on source
counts in the $<$2~keV bands to apply source classification criteria
determined in previous \XMM studies. These ratios were developed
  specifically for use with \XMM data to take advantage of the
  soft-band sensitivity.  They are

\noindent
HR$_{1C}$ = 

\noindent
(SCTS$_{0.5-1.0\ {\rm
      keV}}-$SCTS$_{0.2-0.5\ {\rm keV}}$)/SCTS$_{0.2-1.0\ {\rm keV}}$

\noindent
and HR$_{2C}$ = 

\noindent
(SCTS$_{1.0-2.0\ {\rm keV}}-$SCTS$_{0.5-1.0\ {\rm
      keV}}$)/SCTS$_{0.5-2.0\ {\rm keV}}$.  
\noindent
We include only sources
with enough counts in these bands to have uncertaintes on HR$_{1C}$
and HR$_{2C}$ of $\leq$0.2. We outline a box that isolates SNRs well,
as previously noted by \citet{pietsch2004} and verified here by the
relative isolation of the \citet{LongSNR} SNRs.  Within this box
  are 89 sources (orange dots) not previously known to be SNRs or
  foreground stars.  We visually inspected these source locations in
  the optical emission line and broadband images from the Local Group
  Galaxy Survey \citep[LGGS,][]{massey2006}.  Those sources
corresponding to bright stars are classified as ``fgStar'' in the
catalog.  Those sources corresponding to extended H$\alpha$ shells are
marked as ``SNR'' in the catalog.  There were 23 foreground stars that
were previously unclassified, and 4 SNRs that were previously unknown,
including one of the brightest X-ray SNRs in M33 (Source 383). All of
these are included in the classifications in
Table~\ref{catalog}. Sixty-two of the 89 sources corresponding to
orange dots remained unclassified after inspecting the optical data.

\subsection{Catalog Matching}\label{matching}

We matched our sources with the catalogs of M06 and T11 using the {\it
  IDL} software package {\it match\_xy} in the Tools for ACIS Review
and Analysis {\it TARA} package \citep{Broos2007}.  This package
determines the most likely match for each source in each catalog, and
also reports sources that were not matched.  Each pair of sources was
tested for spatial coincidence, {\it i.e.} that the sources are
  random samples drawn from Gaussians with identical means and
  variances. For the variances we used the positional errors on the
  sources in each catalog. The algorithm also returns the relative
  shift between the catalogs that maximized the number of source
  matches. In addition it has the advantage of providing visualization
  of matches via ds9 region files, forcing one-to-one matching, and
  allowing individual position errors to be specified
  \citep{Broos2010}.  It has been used by \citet{Broos2011} to match
  the {\it Chandra} Carina Complex Project (CCCP) catalog sources to
  their counterparts. Failures in matching result in spurious matches
  and missed matches, which we address below.

Three columns in our catalog (Table 3) show the matches: two columns
(10 and 11) give T11 and MO6 matches, while column 12 gives a
secondary match for the 5 sources that each correspond to a blend of
two T11 sources and the single M06 blend.  In comparing positions for
the matched sources, we found that the T11 positions were more precise
than ours and that we could improve our absolute astrometry by
slightly shifting the positions of the entire catalog by $+$0.1$''$ in
RA and $+$0.7$''$ in Dec.  After applying these shifts, the
  median difference between the RA and Dec of the matched sources is
  zero, with an associated 1$\sigma$ uncertainty of 0.07$''$ in both
  RA and Dec.  The RMS of the matched sources was 2.2$''$ and 1.7$''$
  in RA and Dec (respectively) prior to shifting, and 2.2$''$ and
  1.6$''$ in RA and Dec (respectively) after shifting.  Our original
boresight corrections (see Section~\ref{alignment}) were determined
using positions measured on each individual observation.  These
initial corrections greatly improved the alignment of all of our
observations.  However, our finished catalog includes all of our
  data, has many more sources (hundreds instead of dozens) and is
  based on more counts for sources in areas where multiple
  observations overlap, thus enabling this improved systematic
  correction.

The positional uncertainties from {\tt emldetect} resulted in many
unmatched sources that were clearly re-detections of T11 sources.  For
these sources to be properly matched we continued to add systematic
error to our catalog's positional errors until we matched all
``close'' unmatched sources to the counterpart found by eye but
originally missed by the algorithm. Each time more systematic error
was introduced, we checked the validity of the new matches and made
certain that no obvious spurious matches were being introduced. We
found that the matching routine performed best when we added a
constant 1.5$''$ to all of our position uncertainties.  This result
suggests that the uncertainties from {\tt emldetect} can often be
underestimated, especially when simultaneously measuring multiple
observations with offset aimpoints.  This extra uncertainty is
included in our final catalog (Table~\ref{catalog}).

Our well-tested matching routine from \citet{Broos2007} also allowed
us to look for unmatched sources in regions that overlapped between
surveys.  This included matching our full catalog with the full
catalog of M06, and with that of T11 for the area of our catalog that
was inside the T11 footprint.  Figure~\ref{unmatched} shows the number
of unmatched sources as a function of signal-to-noise of the
detection.

Of the 1296 sources in our catalog, 810 were unmatched to either the
T11 or M06 survey.  Of the 662 {\it Chandra} sources in the T11
catalog, 348 were matched to 343 of our \XMM sources. The other 314
were not detected by our observations.  Inside of the T11 footprint,
our catalog contains 280 sources that did not match any sources in
T11.  All of these sources have fluxes that were above the {\it
  Chandra} sensitivity limit during our observations; however, only
207 of these are not flagged as possible artifacts in our catalog. Of
the 350 sources in the M06 catalog, 306 were matched to 305 of our
sources.  Of these matched sources from our catalog, 162 were also
matched to T11 sources.  Thus 810 of our sources are new detections
not in either of these previous surveys.  Of these 810, 268 are inside
of the T11 footprint, and 195 of these 268 are not flagged as possible
artifacts.  The other 542 unmatched sources are outside of the T11
footprint, and 425 of these are not flagged as possible artifacts.


\subsection{Long-term Variability and Transients}

With our catalog matched to previous surveys of M33, we searched for
transient sources which were not detected in at least one survey where
they should have been above the detection limit.  In addition, we were
able to assess long-term source variability through comparison of
fluxes between catalogs.  We found 51 sources that varied with
5$\sigma$ significance between surveys and 21 sources detected only by
our survey that were bright enough to require a factor of 10 change in
brightness, for a total of 72 sources in our catalog with significant
long-term variability.

\begin{figure}
\centerline{\epsfig{file=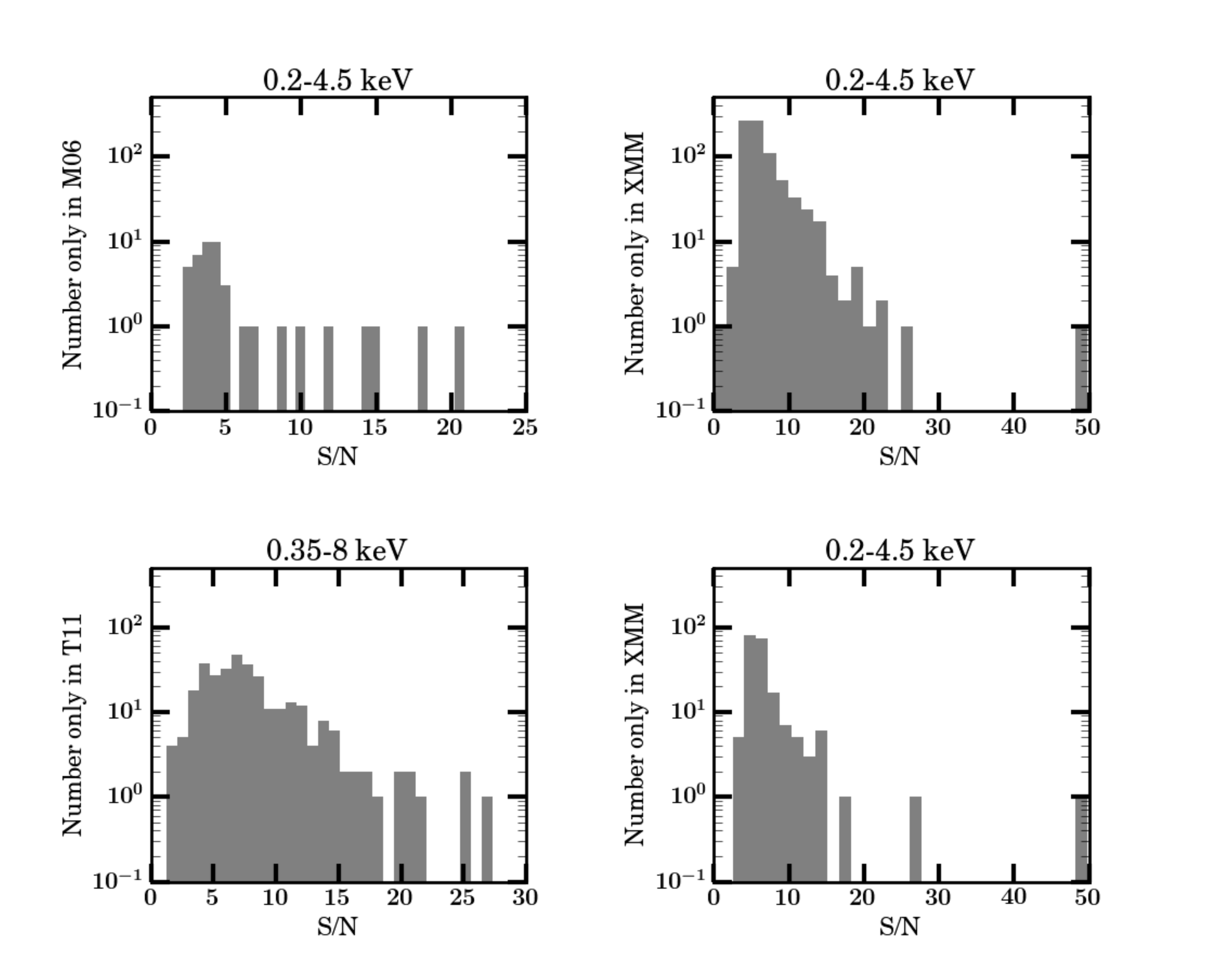,width=6.5in,angle=0}}
\caption{Unmatched sources as a function of signal-to-noise of the
  detection in the 3 matched catalogs (M06, T11, this work).  {\it
  Upper-left} S/N of sources in M06 unmatched to our catalog. {\it
  Upper-right} S/N of sources in our catalog unmatched to M06. {\it
  Lower-left} S/N of sources in T11 unmatched to our catalog. {\it
  Lower-right} S/N of sources in our catalog within the T11 footprint,
  but unmatched to any T11 source.}
\label{unmatched}
\end{figure}

The top panels of Figure~\ref{unmatched} compare our catalog with M06.
Here, our lower signal-to-noise sources were not detected in their
data; this result was expected, as our data are significantly deeper.
However, there are five sources in our catalog with S/N$>$20 that
  were not seen by M06.  Three of these (our sources 203, 651, 861)
  were matched to T11 sources.  The fourth was our source 712, the
  known pulsar transient \citep{trudolyubov2013}.  The fifth (our
  source 1022) lay just outside of the M06 observations.  On the other
  hand, eleven M06 sources with S/N$>$5 in their catalog were not seen
  in our observations, which suggests variability by a factor of at
  least 10 in eight cases (M06 sources 41, 96, 134, 149, 180, 207,
  246, and 296).

The bottom panels of Figure~\ref{unmatched} compare our catalog with
T11 inside of the common area.  Here, we expect their data to be
deeper, and we expect not to have detected sources that had low
signal-to-noise in their data.  However, four of the 41 sources with
S/N$>$12 in T11 require variability in flux by a factor $>$10 to explain
our non-detection.  These sources (T11 sources 13, 26, 233, and 283),
along with all other sources that required variations in flux by a
factor $>$10 are provided in Table~\ref{transients}.

Two of the non-detections correspond to transients reported in
\citet[][XRT-1 and XRT-6]{williams2008} and require a change in flux
by a factor $>$10 to explain their non-detections.  Four of the other
transients reported by \citet{williams2008} were detected by our
survey.  These detections are XRT-2 (our Source 511), XRT-4 (our
Source 586, foreground star), XRT-5 (our Source 480), and XRT-7
(our Source 1128).  Our new detections suggest that these faint
transients did not vary by more than a factor of 10 from their previous
detections.  Therefore they are not included in our list of transient
candidates from this survey. Another (XRT-3 ,T11 source 260), fails to
make it into Table~\ref{transients} because its original detection was
so faint that it requires only a factor of 6.5 change in flux to
explain our non-detection.

\begin{figure}
\begin{minipage}[b]{0.5\linewidth} 
\centering\epsfig{file=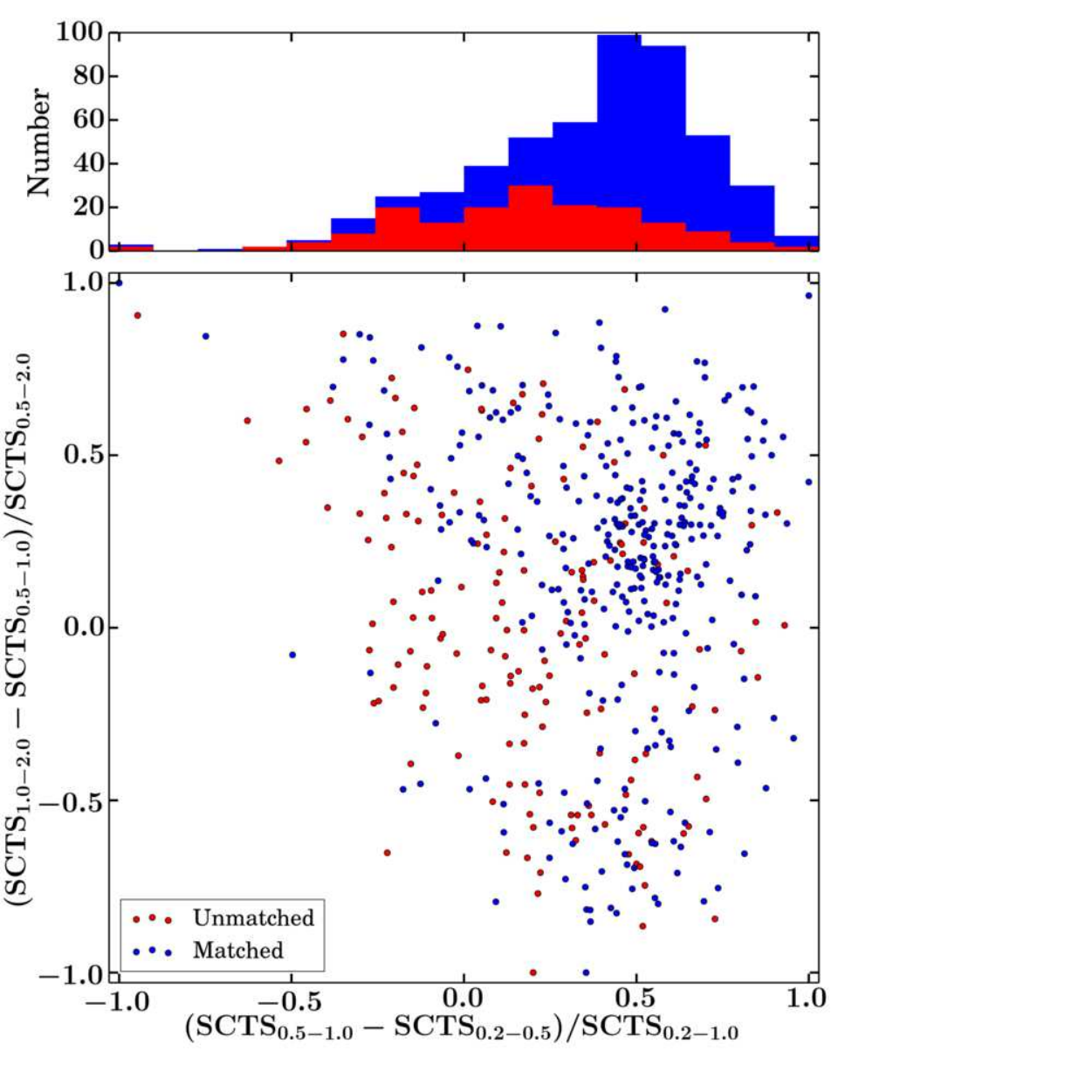,width=3.0in,angle=0}
\end{minipage}  
\mbox{\hspace{0.5cm}}
\begin{minipage}[b]{0.5\linewidth}   
\centering \epsfig{file=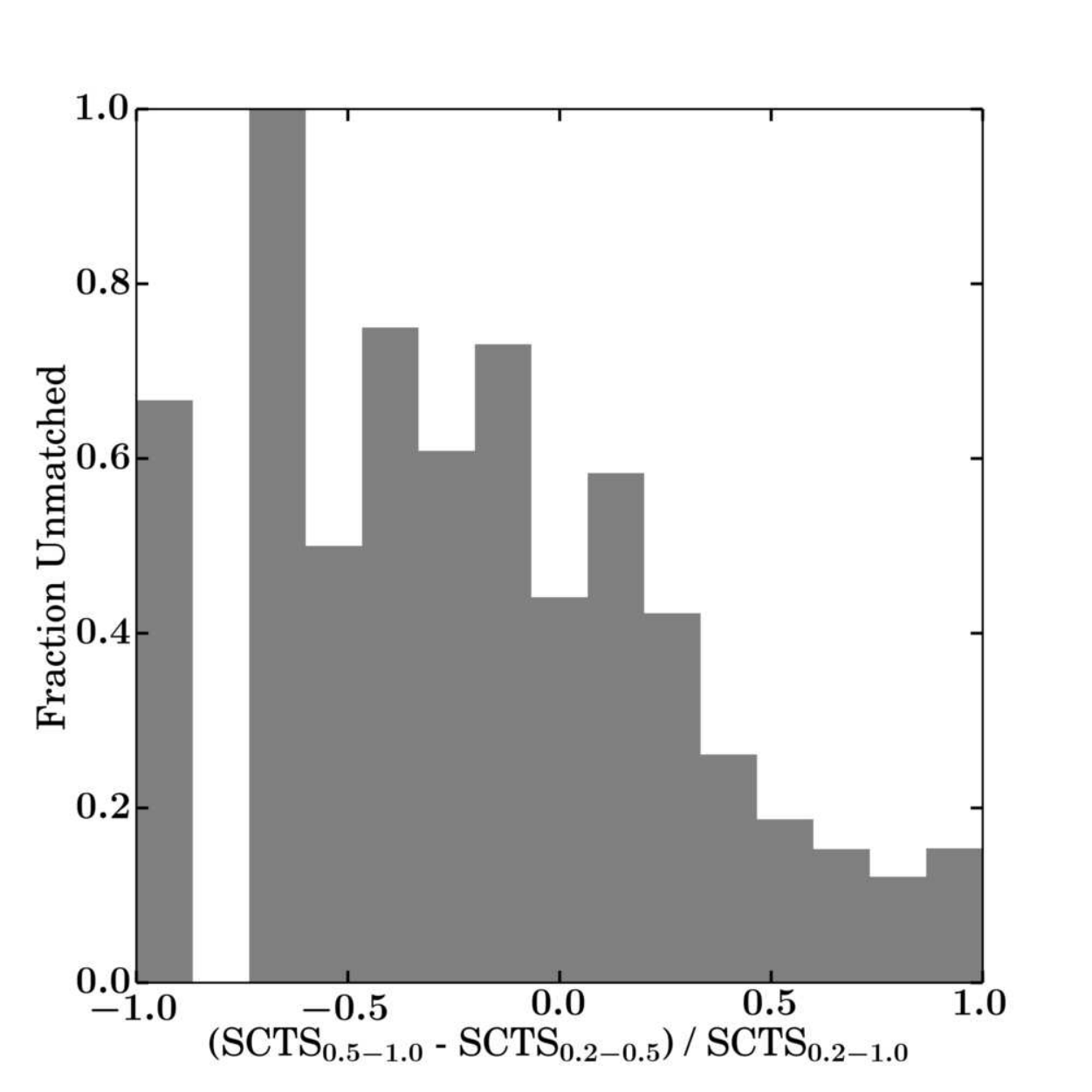,width=3.0in,angle=0}
\end{minipage}  
\caption{ {\it Left Bottom:} Hardness ratios optimized for separating
  soft sources using source counts (SCTS) in the 0.2-0.5, 0.5-1.0, and
  1.0-2.0 keV energy bands of all matched sources inside of the T11
  footprint (blue circles) and unmatched sources with 3$<$S/N$<$8 from
  our catalog inside the T11 footprint (red circles).  SCTS are used
  for the hardness ratios in this case to directly compare the
  relative hardness of the newly-detected sources with the rest of the
  catalog. There are two sources with zero counts in the 0.5-1.0
    keV band, one in the ``unmatched" category and one in the
    ``matched" category. The median uncertainty on the X-axis is 0.16
    for the matched sources and 0.25 for the unmatched sources. The
    median uncertainty on Y-axis is 0.11 for the matched sources and
    0.23 for the unmatched sources. {\it Left Top:} Sandpile
  histogram of all sources shown in {\it left bottom} so that the
  contribution of each to the total in each bin can be compared.  The
  blue area shows the contribution of all matched sources; the red
  shows the contribution of unmatched sources.  The unmatched sources
  have a softer distribution than the overall distribution.  {\it
    Right:} Fraction of unmatched sources as a function of hardness
  ratio, showing that a high fraction of soft sources are not matched
  to a T11 source. }
\label{unmatchedHR}
\end{figure}

Figure~\ref{unmatched} shows that many sources detected by our survey
at S/N$\sim$3--8 and located inside the T11 footprint were not
detected in the {\it Chandra} survey even though they are brighter
than the T11 limiting flux. To investigate the nature of these
sources, we inspected the spatial and hardness distributions of the
unmatched sources.  In Figure~\ref{unmatchedHR} we plot the hardness
ratios based on soft-band source counts (see Section~\ref{hrs_text})
of these 3$<$S/N$<$8 sources unmatched to T11 along with all of the
sources inside the footprint of their survey.  The unmatched sources
are peaked at significantly softer ratios than the matched sources,
indicating that the dominant cause of the discrepancy between the
catalogs is the {\it XMM-Newton} soft response.  Because of this
increased soft sensitivity, our \XMM survey has discovered more faint
soft sources than the {\it Chandra} one (T11).

Because many of the unmatched sources with S/N$\lap$8 in our catalog
appear to be due to increased soft sensitivity and not variability, we
concentrated on separating out the most highly variable transients,
which appear to be those with S/N$>$8 in our catalog.  There were 31
of our catalog sources inside the T11 footprint that were not seen by
T11, but are detected at S/N$>$8 in our data.  The upper limits, taken
from the sensitivity of T11 (see Table~\ref{transients}), suggest
variability by a factor of 10.  Of these, 21 are included in the
table, with the other 10 omitted because we have classified them as
probable SNRs.  SNRs are not variable; instead their non-detections in
T11 arise from the lower soft sensitivity of {\it Chandra}. Comparing
these same 31 sources against M06, we find that 14 of the 21 non-SNR
sources were also not seen in M06.  One (our Source 712) is the
transient pulsar reported by \citet{trudolyubov2013}.

We provide classifications for transients that are likely foreground
stars in the Type column of Table~\ref{catalog}.  Most of these
sources were previously classified by M06 or T11.  However, some are
new classifications based on comparison of soft sources with optical
images from the Local Group Galaxy Survey  (see
  \S~\ref{hrs_text}).

Another interesting previously known variable source in our catalog is
the second eclipsing high mass X-ray binary [PMH2004] 47 (our source
234), which was the target of the PMH observations (Table 1). During
these monitoring observations the source was detected four times
(observations 0606370901, 1001, 1101 and 1201 distributed over 10
days) at a similar flux level of
$\sim$3$\times$10$^{-14}$~erg~cm$^{-2}$~s$^{-1}$. During the other
observations the source was not detected and upper limits indicate
another factor of 10 lower flux.  This is almost a factor of 10 below
the maximum flux reported by \citet{pietsch2009}. Similarly, the mean
flux for the source in the T11 catalog is much higher than in our
catalog, confirming its long-term variability between the surveys.

As a final variability check, we compared the fluxes for all our
sources with a match in T11. The results are provided in
Table~\ref{matchedt11}.  We fit an empirical relation between our
fluxes and those in the T11 catalog in the 0.5-2 keV band, and then
renormalized their fluxes to account for any systematic offset between
our flux measurements. We calculated the variability significance
between the two matched sources before accounting for any systematic
offset ($\sigma$), as well as the variability significance after
accounting for a systematic offset (revised $\sigma$).

\subsection{Short-term Variability}\label{newtransient}

\begin{figure}
\epsfig{file=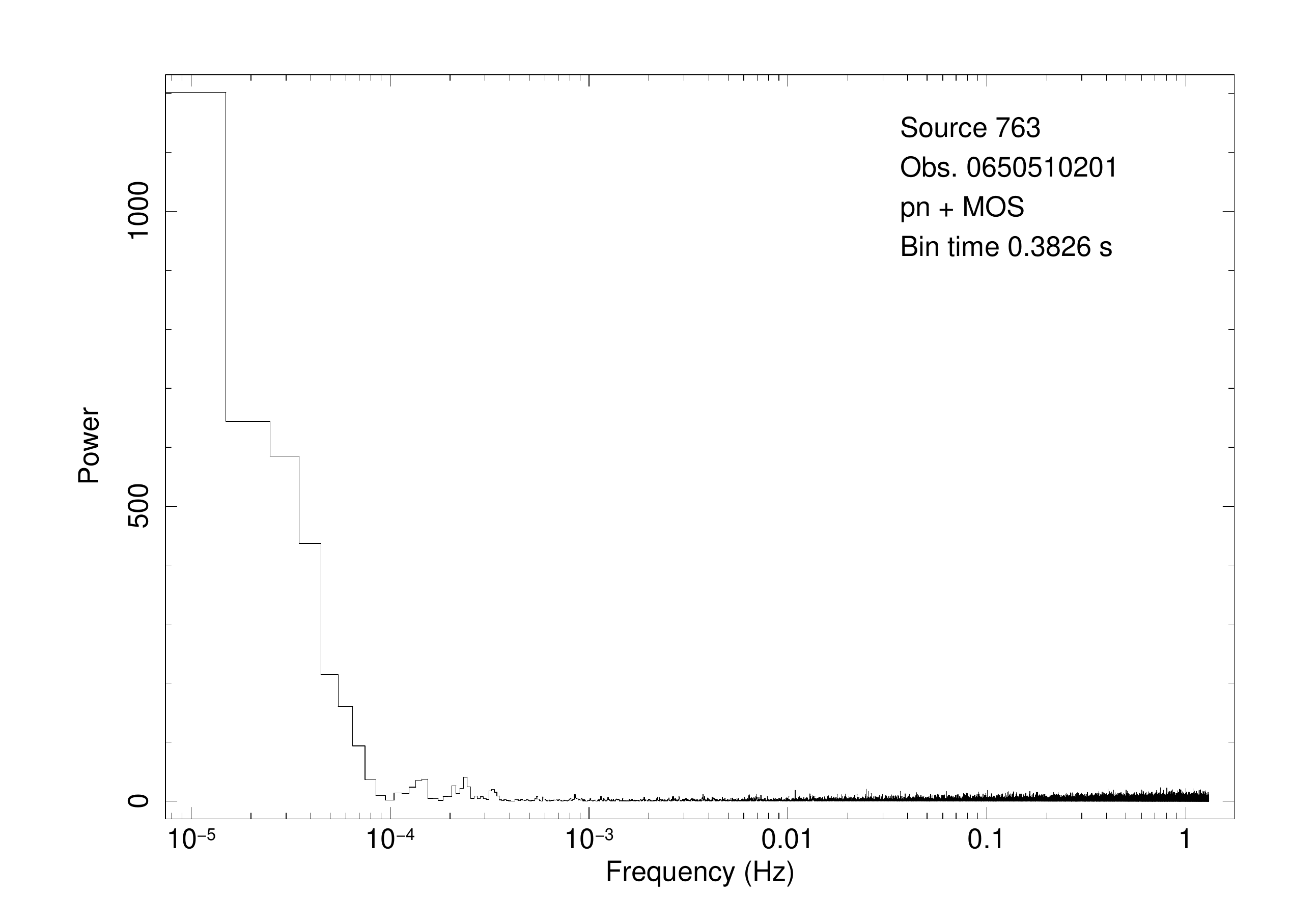,height=2.5in,angle=0}
{\hbox{\hspace{-4.7cm}\centerline{\epsfig{file=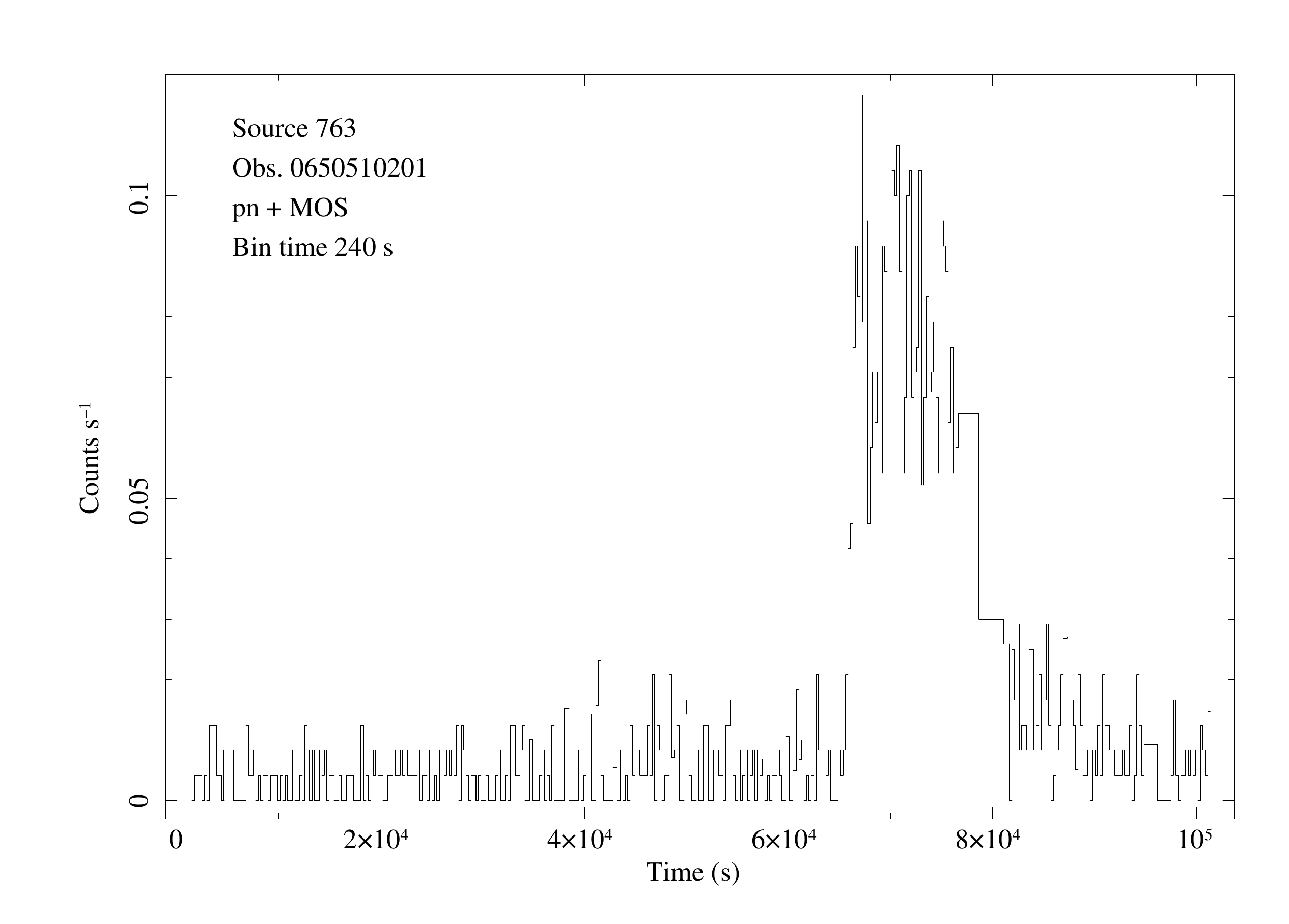,height=2.5in,angle=0}}}}
\epsfig{file=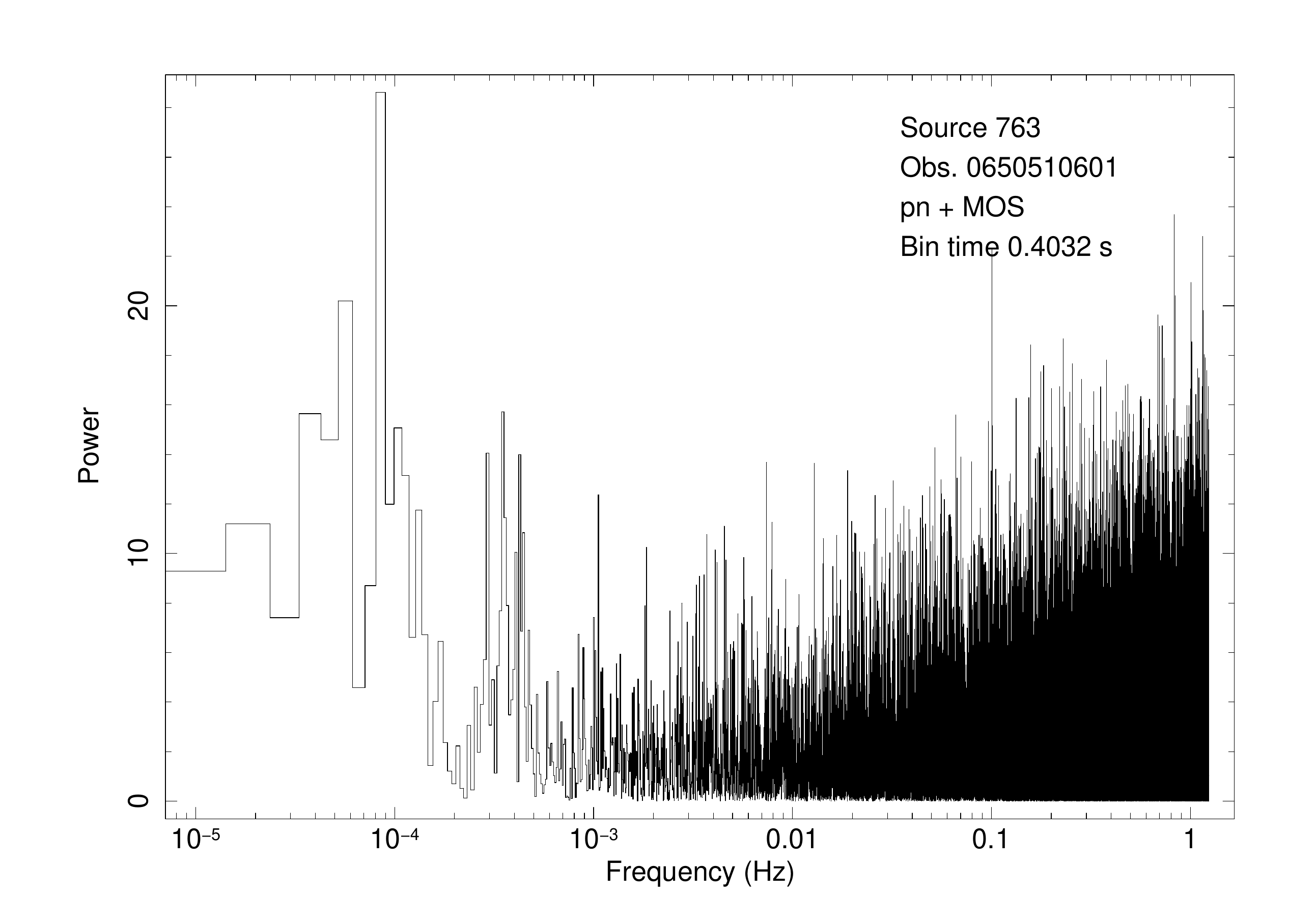,height=2.5in,angle=0}
\epsfig{file=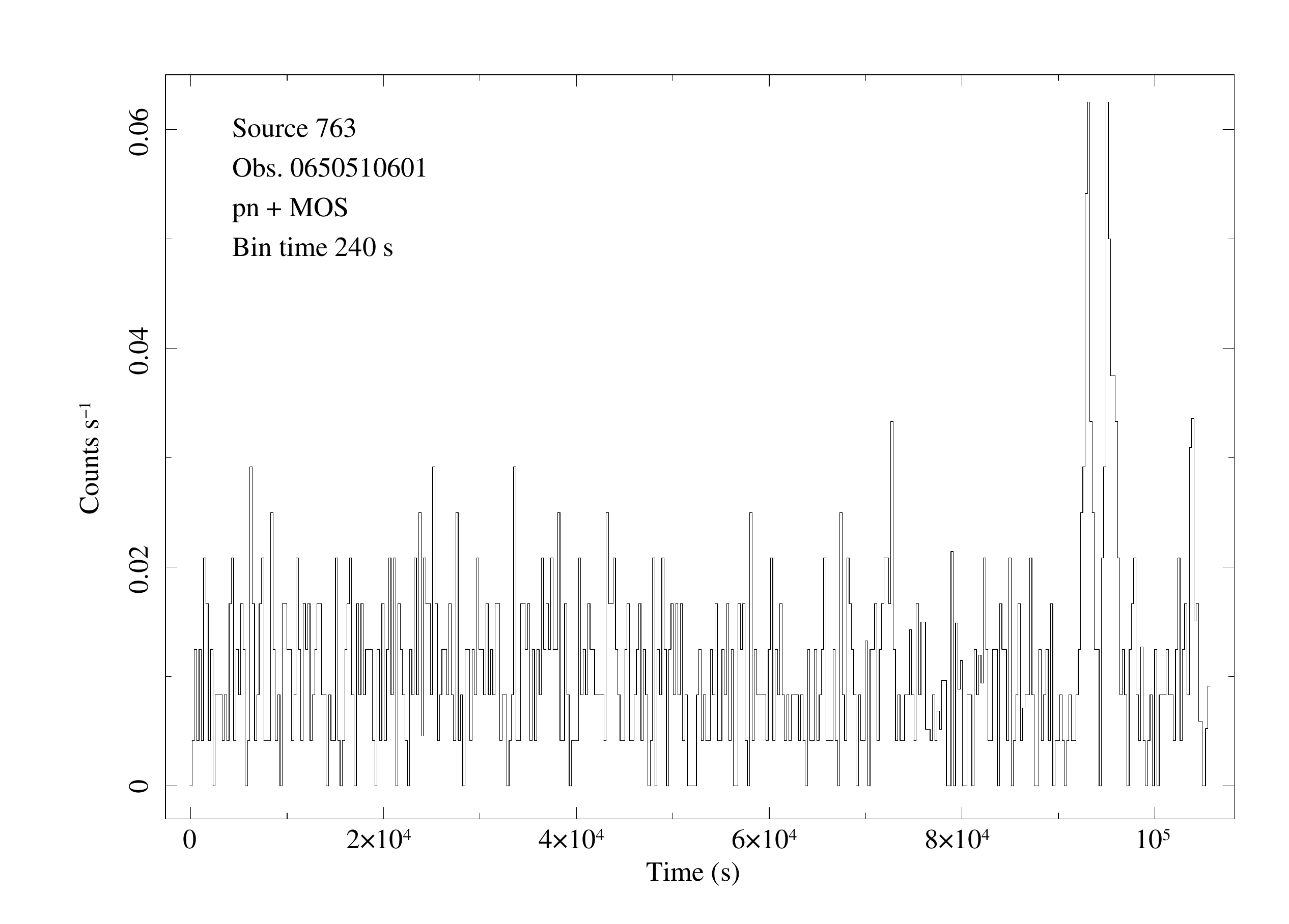,height=2.5in,angle=0}
\caption{Power spectra and lightcurves for two examples of
  observations of Source 763. {\it Top Left:} Power spectrum from
  observation 0650510201, showing a clear detection of variability at
  low frequencies in the power spectrum.  {\it Top Right:} The
  variability suspected by inspecting the power spectrum is confirmed
  with a variable lightcurve. {\it Bottom Left:} Same as {\it top
    left}, but for observation 0650510601, which is a more borderline
  case.  Here variability is seen at much lower significance at low
  frequencies in the power spectrum.  {\it Bottom Right:} Same as {\it
    top right}, but for observation 0650510601.  Again the detection
  is weaker, but variability is confirmed in the lightcurve. }
\label{power}
\end{figure}

In addition to the 72 sources showing long-term variability, we found
15 sources that exhibited variability on the timescale of a single
observation, two of which (our Sources 712 and 783) were also
long-term variables.  We computed power spectra for all of the sources
in our catalog with more than 200 counts in individual
observations. For each observation, power spectra were created for the
EPIC instruments separately and combined (using common GTIs). From
visual inspection of the power spectra, we found 15 sources with
excess power at low-frequency or with a strong peak.  Examples are
shown in Figure~\ref{power}.  These variable sources were then
confirmed through inspection of the lightcurves and are detailed in
Table~\ref{variables}. The power spectra of most of these lightcurves
are characterized by excess power at low frequencies caused by flares,
in support of their identification with foreground stars. One source
(our Source 712) was the pulsar reported by \citet{trudolyubov2013},
which was visible in the observations of Fields 1 and 2, and another
(our Source 521) was the known eclipsing HMXB X-7 \citep{pietsch2006}.
Thus, our data were sensitive enough to detect such periodic
variability, and it appears that M33 contains such sources.

Another short-term variable (our Source 128) was an interesting
transient that appeared in only a single, short observation.  Although
this source was too faint on average to show that it varied by a
factor of 10 across surveys (and therefore is not included in
Table~\ref{transients}), we measured it to be transient within the PMH
observations. This transient candidate (RA$=$01:32:15.06,
Dec$=$+30:32:21), is near a red star at RA$=$01:32:15.16,
Dec$=$+30:32:23.4 with B=21.04 mag and V=19.07 mag in the
  Johnson system.  The lightcurve does not look like a typical
foreground flare, and the star is 4$\sigma$ away from the source
position.

We extracted the source spectrum within a 19$''$ radius circle and a
background spectrum within a 50$''$ radius circle nearby on the same
EPIC-pn CCD; we included single and double pixel events, with FLAG=0.
The spectrum was binned to have a minimum of 20 cts/bin.  All
single-component models (thermal bremstrahlung, power-law,
disk-blackbody, and blackbody) produced unacceptable fits
($\chi^2/dof>21/14$).  A disk-blackbody with a power-law produces a
very good fit with $\chi^2/dof=13.8/12$.  The power-law index is
2.1$\pm$1.6, and the disk-blackbody temperature is 0.16$\pm$0.06.
Although the source could be a transient LMXB, the low temperature and
possible optical counterpart allow the possibility that it was a flare
from a background AGN.  We were not able to determine a reliable
classification for this transient source, and this position is clearly
of interest in any future observations.

As detailed in Table~\ref{variables}, the other sources with
significant short-term variability were all flares and coincident with
red point sources consistent with foreground stars.

\section{Population Characteristics}

With our broad coverage and depth we were able to measure the source
surface density out to D$_{25}$ and beyond, and we were able to
measure the XLF accounting for the local background explicitly.  These
measurements are described below.

\subsection{Radial Source Density}

\begin{figure}
\centerline{\epsfig{file=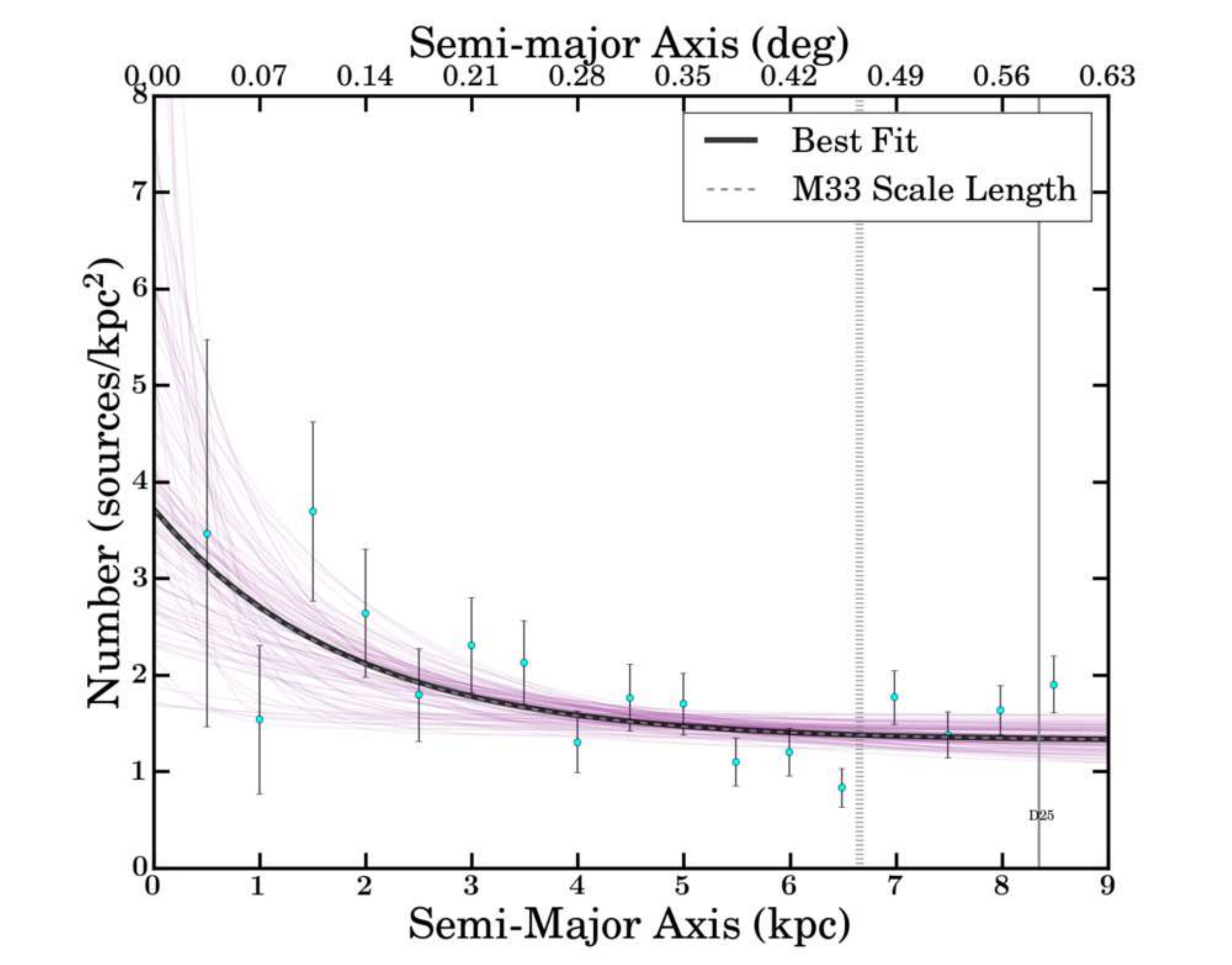,width=6.5in,angle=0}}
\caption{X-ray source surface density for all sources with
  L$_{0.2-4.5}{>}3.6{\times}10^{35}$ erg~s$^{-1}$ and flags described
  in Section~4.2. Thick black line shows an exponential with scale
  length 1.8 kpc (best fit) and constant background of 1.3 kpc$^{-2}$.
  For comparison, the optical scale length (1.8 kpc) is plotted with a
  gray dashed line, which lies on top of the black line. A random draw
  of 100 trials from the MCMC runs used for uncertainty determination
  is shown with the thin pink lines.  A constant surface density,
  which we take to be the background surface density, is reached by
  6.7 kpc (marked with the vertical dotted line) and is relatively
  well-constrained, while the scale length is less certain due to the
  large error bars in the inner disk. D$_{25}$ is marked with the
  vertical gray line.}
\label{radial}
\end{figure}

Many of the sources in our catalog are probably background AGN.  In
order to study the properties of sources in M33 itself, we
investigated the radial density for the bright sources.  We divided
our catalog into elliptical annuli assuming an M33 position angle of
23$^{\circ}$ and an inclination of 54$^{\circ}$
\citep{devaucouleurs1991}.  Figure~\ref{radial} shows the source
density as a function of semi-major axis of the ellipse (radius
equivalent).  We include only sources with 0.2-4.5 keV fluxes brighter
than 4.5$\times$10$^{-15}$~erg cm$^{-2}$ s$^{-1}$
(L$_X{>}$3.6$\times$10$^{35}$~erg s$^{-1}$), as faintward of this
limit the catalog becomes increasingly background-dominated.  This cut
includes the 391 brightest sources in the catalog.  We include sources
with the ``s'' and ``t'' flags in this analysis, and sources flagged
as possible artifacts with a match to T11 (as matches to T11 are
likely real sources).

We fitted this radial profile with an exponential plus a constant
($\Sigma_{N_X}=Ae^{-r/r_s}+B$, where $\Sigma_{N_X}$ is the source
density, $A$ is the normalization of the exponential term, $r$ is the
galactocentric radius, $r_s$ is the exponential scale length, and $B$
is the constant background density).  We assume the exponential term
of the function represents the M33 population while the added constant
term represents the AGN background.  The fit was performed using a
Markov Chain Monte Carlo technique, and specifically the {\it emcee}
Python module \citep{foreman-mackey2013}. A random draw of 100 samples
from the fitting used for uncertainty determinations are shown with
the thin lines on the figure. The errors on data points are
Poisson. 

We find the source density falls with an exponential scale length
($r_s$) of 1.8$^{+1.3}_{-0.8}$ kpc, indistinguishable from the optical
scale length of the disk \citep[$\sim$1.8 kpc;][]{williams2009b}. The
normalization ($A$) of the exponential source density term was
2.3$^{+1.6}_{-1.0}$ sources kpc$^{-2}$.  Integrating the M33 component
(the exponential term) out to D$_{25}$ yields 50$^{+40}_{-30}$ M33
sources down to 3.6$\times$10$^{35}$~erg~s$^{-1}$.  Since there are 20
known SNRs at these luminosities \citep{LongSNR}, this analysis
suggests that roughly 40\% of the bright X-ray sources in M33 are
SNRs.  However, we note that this number of M33 sources is likely
biased low, as we do not account for the loss of background sources
due to absorption by M33 itself. Absorption effects would decrease the
background contribution within D$_{25}$, thus increasing the number of
M33 sources.

Our profile fit also constrains the background source density
(the constant added to the exponential) to
460$^{+40}_{-70}$~deg$^{-2}$ ($B$=1.3$^{+0.1}_{-0.2}$~kpc$^{-2}$, as
in Figure~\ref{radial}, inclination-corrected), roughly consistent
with what would be inferred from the \citet{cappelluti2009}
  densities at these flux levels, yielding a total of
290$^{+30}_{-40}$ background sources inside of D$_{25}$ (0.59x0.37
deg, 8.4 kpc deprojected).

\subsection{Luminosity Function}

\begin{figure}
\centerline{\epsfig{file=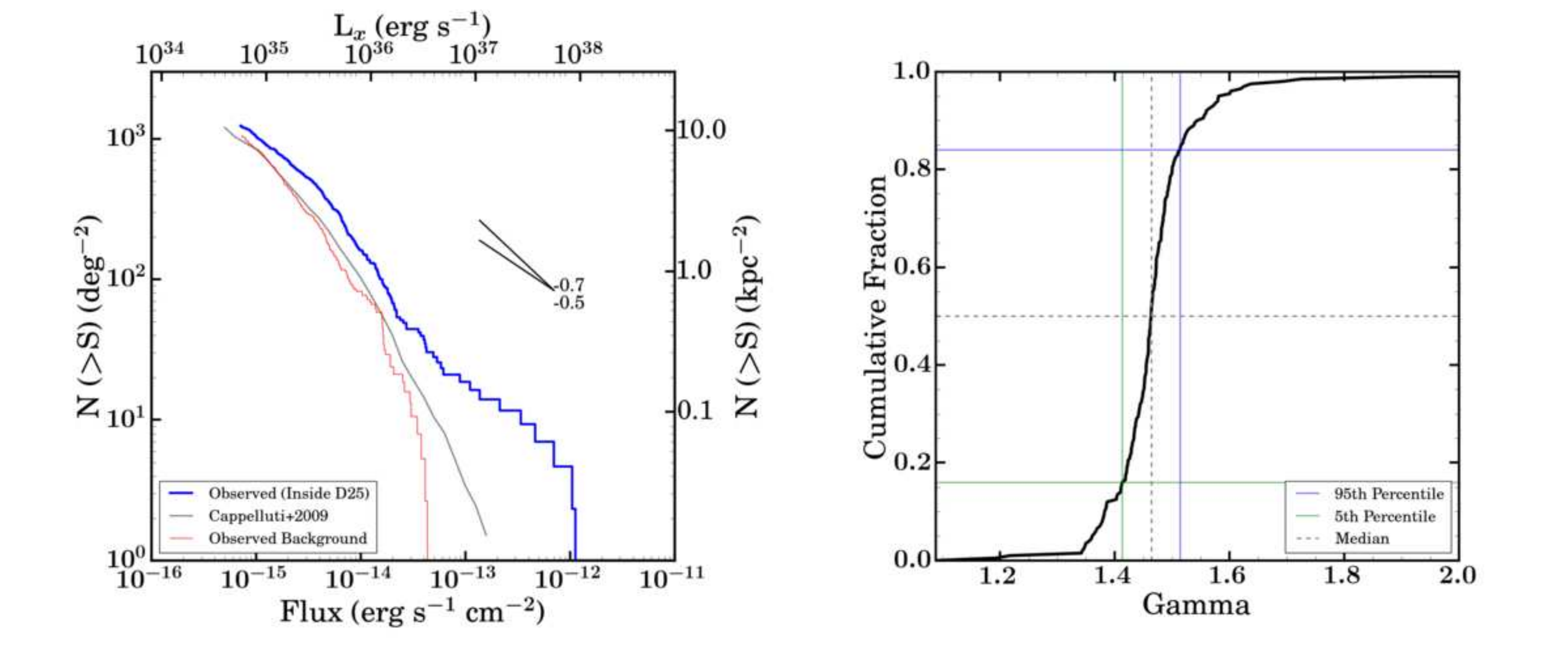,width=6.5in,angle=0}}
\caption{{\it Left:} Cumulative luminosity function of all sources
  with fluxes $>$6$\times$10$^{-16}$ erg~cm$^{-2}$~s$^{-1}$ inside of
  6.7 kpc (deprojected) is shown in blue.  Cumulative power-law
  indices of 0.5 and 0.7 (equivalent to 1.5 and 1.7 differential
  indices) are shown for comparison.  The cosmic mean background
    shown in gray is taken from \citet{cappelluti2009}. The observed
    background, shown in red, is measured from our data outside of 6.7
    kpc. {\it Right:} Cumulative distribution of the best fit power
    law index of the M33 component of the XLF from fits to our Monte
    Carlo draws from the catalog fluxes.  The resulting 90\%
    uncertainties to the best fit are 1.50$^{+0.08}_{-0.14}$.}
\label{XLF}
\end{figure}

\begin{figure}
\centerline{\epsfig{file=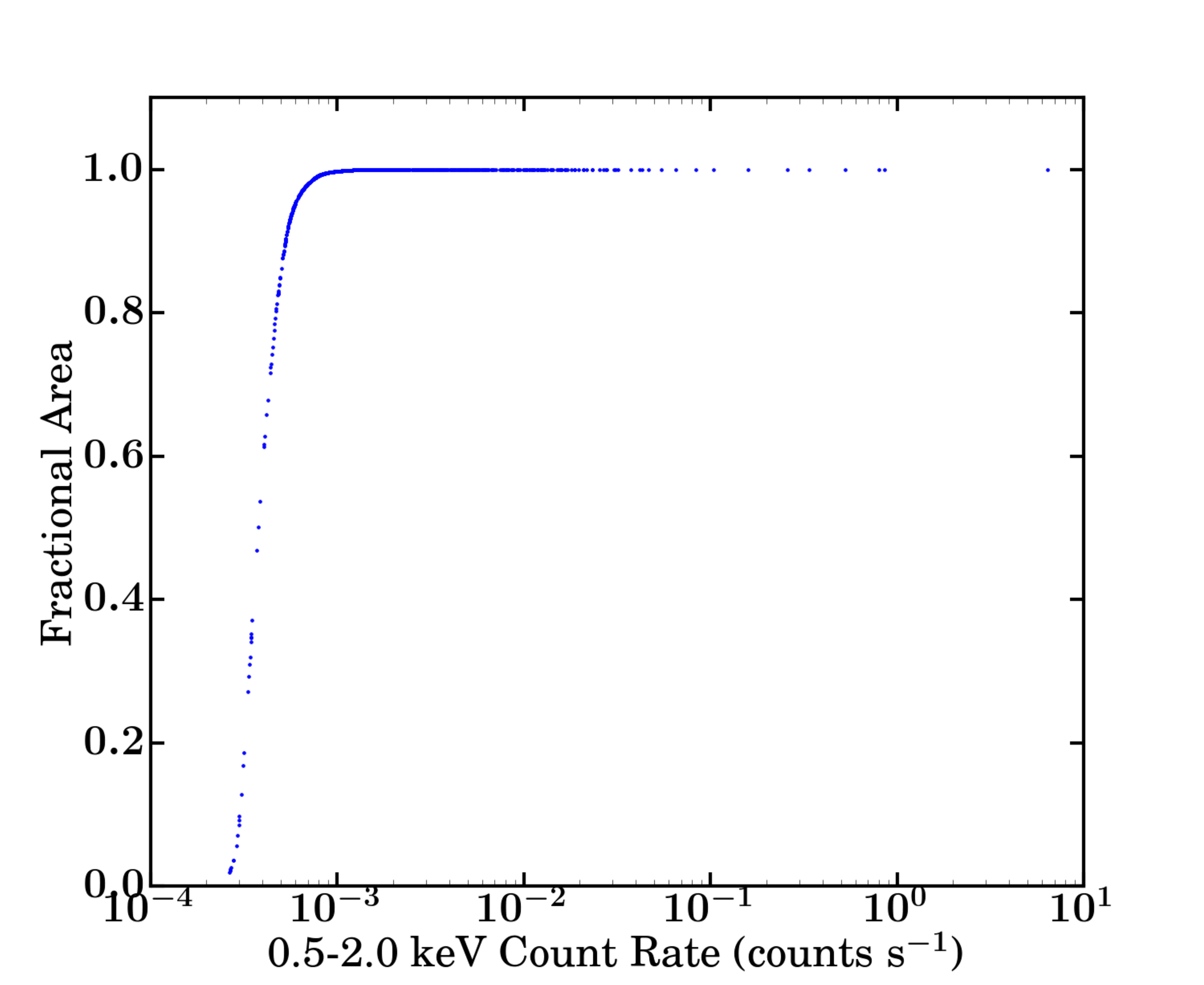,width=4.5in,angle=0}}
\caption{Our fractional coverage of the area inside 6.7 kpc
  (deprojected) as a function of 0.5-2.0 keV count rate, as calculated
  from our combined sensitivity map.  The entire area is covered down
  to a count rate of $\sim$8$\times$10$^{-4}$ counts s$^{-1}$.}
\label{sensitivity}
\end{figure}

In Figure~\ref{XLF} we plot the XLF (log(N)-log(S)) of our survey
within 6.7 kpc (deprojected) of the center of M33 (just inside of D25,
as marked on Figure~\ref{radial}), where N has units of deg$^{-2}$ and
L is the 0.5-2.0 keV luminosity of the source assuming all sources are
at the distance of M33.  This XLF excludes sources classified as
  foreground stars, and includes all unflagged sources, sources with
  the ``s'' or ``t'' flags, and otherwise flagged sources only if
  matched to a T11 source.  For those sources with the ``s'' flag, we
  take the 0.5-2.0 keV fluxes from Table~\ref{catalog} (see
  description of Column 9 in Section~\ref{cattext}).  The ``t'' flag
  does not affect our XLF measurement, as we are not using the total
  0.2-4.5 keV fluxes here.  Each source was corrected for completeness
  by calculating the amount of our survey area inside of 6.7 kpc
  sensitive to the source's count rate.  We show the fractional area
  sensitive to each count rate in Figure~\ref{sensitivity}.  The
  lowest 0.5-2.0~keV count rate on the area sensitivity plot
  (2.7$\times$10$^{-4}$) corresponds to a 0.5-2.0~keV flux of
  $\sim$2$\times$10$^{-16}$ erg cm$^{-2}$ s$^{-1}$. The luminosity
  function itself is limited to sources with 0.5-2 keV fluxes above
  6$\times$10$^{-16}$ erg cm$^{-2}$ s$^{-1}$, or 5$\times$10$^{34}$
  erg s$^{-1}$ ($\sim$8$\times$10$^{-4}$ combined 0.5-2.0 keV count
  rate) for which the completeness corrections are small.  Thus, our
  area corrections are not sensitive to the details of the sensitivity
  map; however, we do correct for completeness using the area
  corrections derived from our sensitivity function.

Figure~\ref{XLF} also shows our measured background XLF.  To find the
observed background we constructed an XLF from all sources outside of
6.7 kpc (the radius showing enhanced surface density, 0.47 deg along
the major axis, marked on Figure~\ref{radial}), but within a region
for which there was at least 100~ks of exposure time.  We corrected
for completeness using the same technique as for the inner regions,
but with areas derived from our sensitivity map in this outer region
of the survey. We also plot the background XLF as determined from
\citet{cappelluti2009}.  The agreement gives us confidence that our
catalog is clean and our completeness function is accurate.

We first fit the unbinned differential XLF with a sum of two
  power-law components using the {\tt CIAO} package {\it Sherpa}.
  First, we measure the background component alone with a single power
  law using only sources measured outside of the surface density
  enhancement seen in the radial analysis.  Then we fit the total XLF
  inside of the surface density enhancement by adding a second power
  law component to the fixed background component.  The second
  power-law represents the intrinsic M33 XLF.

The best-fit to the background XLF (323 sources outside of the area of
enhanced surface density in the radial analysis) has a power-law index
of 1.83.  To assess the precision of this background power-law index,
we performed 200 Monte Carlo draws from the fluxes, where each source
was given a flux drawn from a Gaussian distribution centered at the
measured flux with a width determined from the Poisson uncertainties.
The 90\% uncertainties from this method are
1.83$^{+0.09}_{-0.05}$. We note that this background XLF
  represents the maximum expected background contamination in the M33
  catalog, because some of the sources {\it directly} behind M33 are
  likely lost because of absorption by M33 itself.

We then added a second power-law component to the model and fit the
unbinned differential XLF for 523 sources inside of the radial surface
density enhancement.  The additional component, which we attribute to
the X-ray source population of M33, has a power-law index of 1.50.  To
estimate the uncertainties on this result, we again performed 200
Monte Carlo draws from our measured source fluxes, using the same
technique applied to the background fits.  However, in addition to
applying uncertainties to the fluxes, for each draw, we also fixed the
background component to a random draw from our 200 background fits.
Thus we also account for uncertainty in the background XLF.  In the
right panel of Figure~\ref{XLF} we show the cumulative distribution of
the resulting power-law index values for the M33 component from these
200 fits.  With the resulting 90\% uncertainties, we measure the
power-law index of the intrinsic M33 XLF to be 1.50$^{+0.08}_{-0.14}$.
If we integrate the M33 component, we find 60$^{+50}_{-30}$ M33
sources, consistent with the radial distribution analysis.  As with
the result of the radial analysis, this value is likely to be biased
low because some background sources are likely lost to absorption by
M33 itself.  This XLF slope is consistent with that measured by T11
and similar to the ``universal'' XLF of HMXBs seen in several external
galaxies \citep[1.6; e.g.,][]{grimm2003,mineo2012}, suggesting that
the bright X-ray source population of M33 has a large fraction of
HMXBs.


\section{Conclusions}

We have carried out a deep {\it XMM-Newton} Survey of M33 to
complement that performed by {\it Chandra}. Our new data provide
increased sensitivity to soft sources and cover the entire D$_{25}$
isophote to similar depth to that probed by {\it Chandra} for the
inner 15$'$ of the galaxy (T11).  

The primary purpose of this paper is to describe the methods we have
used to produce a catalog of the point sources contained in the \XMM
survey.  We have described new methods for reducing and analyzing overlapping
observations with {\it XMM-Newton} to produce catalogs with
well-measured source properties.  These techniques take full advantage
of the extra depth in the overlapping regions and minimize ambiguity
associated with post-facto combining of separately-measured catalogs
for each observation.

Our final catalog contains 1296 sources.  Our complete coverage and
soft sensitivity have resulted in 810 new source detections, 620 of
which are not flagged as possible artifacts (see Column 9 description
in Section~\ref{cattext}).  We find that many of the soft sources in
our catalog were previously undetected, highlighting the value of the
soft sensitivity of {\it XMM-Newton}. 

Furthermore, the depth and coverage have allowed an extended radial
profile of the M33 X-ray source density out to D$_{25}$ and beyond,
which has a scale length consistent with the optical scale length,
similar to other nearby spiral galaxies \citep[e.g.][]{binder2012}.
Our radial profile suggests that a relatively low number of the
sources ($\sim$50, 15\%) with fluxes $>$4.5$\times$10$^{-15}$ erg
cm$^{-2}$ s$^{-1}$ belong to M33, and about 40\% (20) of these are
known SNRs.  However, this number of M33 members is likely to be
biased low since some background sources may be undetected through
M33.

Finally, our data allow a local characterization of the background
XLF, which we measure to have a differential power-law index of
1.83$^{+0.05}_{-0.09}$.  When we account for this background, we find
the differential XLF of M33 itself has a power-law index of
1.50$^{+0.08}_{-0.14}$, consistent with previous measurements and the
``universal'' XLF for HMXB populations.

Future papers will use this catalog to discuss the detailed properties
of sets of sources.  In particular, a detailed study of the X-ray
spectra of the supernova remnants is in preparation, and we plan to do
more work on point source optical counterpart identifications and
characteristics.

Support for this work was provided by NASA grants NNX12AD42G and
NNX12AI52G. TJG and PPP acknowledge support under NASA contract
NAS8-03060 with the {\it Chandra\/} X-ray Center."




\begin{table}[h!]
\begin{center}
\caption{XMM-Newton Observations used for the survey}
\scalebox{0.65}{
\begin{tabular}{|c|c|c|c|c|c|c|r|r|r|r|r|}
\hline
Field ID & Obs. ID 	& Start Date	& End Date	& RA(J2000.0) 		& Dec(J2000.0)	 	& Roll Angle	& \multicolumn{2}{|c|}{Exposure\tablenotemark{1} (ks)} & \multicolumn{3}{|c|}{Eff Exp. (ks)}\\
	&		&		&		& (J2000)	& (J2000)	& (Deg.)	& MOS &   PN	& MOS1 & MOS2 & PN \\
\hline
1	& 0650510101	& 2010-07-09	& 2010-07-10	& 01:34:10.05	& +30:46:53.3	& 065:00:00.0	& 101.0	& 99.5	& 85.4 & 85.4 & 83.8	\\
2	& 0650510201	& 2010-07-11	& 2010-07-12	& 01:33:42.05	& +30:34:53.3	& 065:00:00.0	& 101.0	& 99.5	& 88.1 & 88.1 & 86.5	\\
3	& 0650510301	& 2010-07-21	& 2010-07-22	& 01:34:41.40	& +31:01:11.8	& 065:00:00.0	& 107.7	& 104.9	& 76.5 & 76.4 & 75.1	\\
4	& 0672190301	& 2012-01-10	& 2012-01-12	& 01:34:55.15	& +30:41:34.5	& 249:58:43.1	& 119.1	& 120.3	& 97.7 & 97.7 & 96.7	\\
5	& 0650510501	& 2010-08-10	& 2010-08-11	& 01:33:19.79	& +30:53:46.9	& 065:00:00.0	& 99.6 	& 88.6	& 74.8 & 74.8 & 67.7	\\
6	& 0650510601	& 2010-08-12	& 2010-08-13	& 01:34:18.33	& +30:26:17.2	& 065:00:00.0	& 127.9	& 124.5	& 103.2 & 103.3 & 101.9	\\
7	& 0650510701	& 2010-08-14	& 2010-08-15	& 01:33:11.13	& +30:19:41.8	& 065:00:00.0	& 99.0 	& 97.7	& 70.1 & 70.1 & 69.0	\\
\hline
PMH-03	& 0606370301	& 2010-01-07	& 2010-01-07	& 01:32:36.90	& +30:32:28.0	& 250:53:41.2	& 14.6 	& 13.0	& 14.5 & 14.5 & 13.0	\\
PMH-04	& 0606370401	& 2010-01-11	& 2010-01-11	& 01:32:36.90	& +30:32:28.0	& 249:28:19.5	& 17.0 	& 15.4	& 15.1 & 15.1 & 13.5	\\
PMH-06	& 0606370601	& 2010-01-17	& 2010-01-18	& 01:32:36.90	& +30:32:28.0	& 247:24:17.6	& 29.4 	& 27.2	& 21.1 & 21.1 & 19.8	\\
PMH-07	& 0606370701	& 2010-01-21	& 2010-01-21	& 01:32:36.90	& +30:32:28.0	& 256:00:00.0	& 15.5 	& 13.9	& 8.2 & 8.2 & 7.0	\\
PMH-09	& 0606370901	& 2010-01-28	& 2010-01-28	& 01:32:36.90	& +30:32:28.0	& 243:21:21.6	& 19.6 	& 18.0	& 17.7 & 17.7 & 16.1	\\
PMH-10	& 0606371001	& 2010-01-31	& 2010-01-31	& 01:32:36.90	& +30:32:28.0	& 225:00:00.0	& 14.6 	& 7.0	& 10.2 & 10.1 & 6.7	\\
PMH-11	& 0606371101	& 2010-02-04	& 2010-02-04	& 01:32:36.90	& +30:32:28.0	& 241:17:33.8	& 14.6 	& 13.0	& 2.2 & 2.2 & 1.7	\\
PMH-12	& 0606371201	& 2010-02-07	& 2010-02-07	& 01:32:36.90	& +30:32:28.0	& 239:52:26.4	& 16.9 	& 15.3	& 15.4 & 15.4 & 15.3	\\
PMH-15	& 0606371501	& 2010-02-24	& 2010-02-24	& 01:32:36.90	& +30:32:28.0	& 244:42:24.1	& 9.5 	& 7.9	& 7.0 & 7.0 & 6.5	\\
\hline
\label{obstable}
\end{tabular}
}	
\tablenotetext{1}{Total available exposure from PMH~47 observations is 166.3 and 143.8 ks for MOS and PN respectively. Total available exposure for the entire survey is 921.6 and 878.8 ks. Good time interval selection provides 111.4, 111.4 and 99.8 ks from the PMH~47 observations for MOS1, MOS2 and PN respectively and 707.2, 707.1 and 680.5 ks total for the entire survey.}
\end{center}
\end{table}


\begin{table}[h!]
\caption{Unabsorbed energy conversion factors (ECF) values for different energy bands and instruments. Units are 10$^{11}$ counts cm$^{2}$ erg$^{-1}$.}
\begin{center}
\begin{tabular}{|c|c|c|c|c|c|c|}	
\hline
Energy Band & MOS1 & MOS2 & PN \\
(keV) & Med Filter & Med Filter & Thin Filter \\
\hline
0.2-0.5  & 0.5009 & 0.4974 & 2.7709 \\
0.5-1.0  & 1.2736 & 1.2808 & 6.006 \\
1.0-2.0  & 1.8664 & 1.8681 & 5.4819 \\
2.0-4.5  & 0.7266 & 0.7307 & 1.9276 \\
\hline
\end{tabular}
\end{center}
\label{ecfs}
\end{table}

\clearpage

\begin{table}
\caption{M33 X-ray Source Catalog: Full version available in
  machine-readable format only.}
\label{catalog}
\scalebox{0.5}{
\begin{tabular}{|ccccccccccccc|}
\hline 
(1) & (2) & (3) & (4) & (5) & (6) & (7) & (8) & (9) & (10) & (11) & (12) & (13)\\ 
Src &  RA(J2000)  &  Dec(J2000)  & r$_\sigma$ & DL & Counts &
Count Rate & Flux & Flag & CXO & XMM & Second & Type\\
  & (h mm ss.ss) & (+dd mm ss.s) & ('') &  &  & (s$^{-1}$) & (erg cm$^{-2}$ s$^{-1}$) & & & & & \\
\hline
351 & 1 33 03.6 & 30 39 03.41 & 1.7 & 1.25e+03 & 1.67e+03$\pm$5.5e+01 & 1.69e-02$\pm$5.6e-04 & 2.85e-14$\pm$1.1e-15 & 0 & 013303.55+303903.8 & 95 & 0 & 0\\
352 & 1 33 03.65 & 30 35 49.64 & 2.6 & 7.02e+00 & 1.01e+02$\pm$2.2e+01 & 1.20e-03$\pm$2.6e-04 & 2.03e-15$\pm$5.1e-16 & 0 & 0 & 0 & 0 & 0\\
353 & 1 33 03.86 & 30 58 56.45 & 3.1 & 2.23e+01 & 7.62e+01$\pm$1.6e+01 & 1.68e-03$\pm$3.4e-04 & 5.23e-16$\pm$1.5e-16 & s & 0 & 0 & 0 & 0\\
354 & 1 33 04.04 & 30 23 03.42 & 2.9 & 4.49e+00 & 1.07e+02$\pm$2.3e+01 & 9.83e-04$\pm$2.1e-04 & 2.12e-15$\pm$4.7e-16 & 0 & 0 & 0 & 0 & 0\\
355 & 1 33 04.08 & 30 39 52.39 & 2.6 & 3.24e+01 & 2.30e+02$\pm$3.0e+01 & 2.45e-03$\pm$3.3e-04 & 3.21e-15$\pm$5.8e-16 & 0 & 013304.03+303953.6 & 0 & 0 & SNR\_L10\\
\hline
\end{tabular}
}
\end{table}

\begin{table}
\scalebox{0.6}{
\begin{tabular}{|cccccccc|}
\hline
\multicolumn{4}{|l}{0.2-0.5 keV Totals}  &  \multicolumn{4}{l|}{0.5-1.0 keV Totals} \\
(14) & (15) & (16) & (17) & (18) & (19) & (20) & (21)\\
DL   &  Counts  &  Count Rate  &  Flux & DL   &  Counts  &  Count Rate  &  Flux \\
    &   &  (s$^{-1}$)  &  (erg cm$^{-2}$ s$^{-1}$) &    &   &  (s$^{-1}$)  &  (erg cm$^{-2}$ s$^{-1}$)  \\
\hline
1.35e+01 & 5.53e+01$\pm$1.3e+01 & 5.52e-04$\pm$1.3e-04 & 1.38e-15$\pm$3.3e-16 & 2.73e+02 & 3.75e+02$\pm$2.6e+01 & 3.74e-03$\pm$2.6e-04 & 4.31e-15$\pm$3.0e-16\\
6.61e+00 & 2.92e+01$\pm$1.1e+01 & 3.48e-04$\pm$1.4e-04 & 7.09e-16$\pm$3.2e-16 & 1.22e-01 & 1.09e+01$\pm$9.5e+00 & 1.31e-04$\pm$1.2e-04 & 1.41e-16$\pm$1.3e-16\\
2.00e+00 & 8.21e+00$\pm$5.3e+00 & 1.71e-04$\pm$1.1e-04 & 4.85e-16$\pm$3.4e-16 & 2.91e+00 & 1.34e+01$\pm$7.6e+00 & 3.00e-04$\pm$1.7e-04 & 3.51e-16$\pm$2.2e-16\\
1.69e-01 & 1.23e+01$\pm$9.0e+00 & 1.06e-04$\pm$7.9e-05 & 1.82e-16$\pm$1.8e-16 & 1.69e-01 & 6.17e+00$\pm$7.6e+00 & 5.43e-05$\pm$6.8e-05 & 1.95e-17$\pm$7.0e-17\\
1.29e+00 & 1.52e+01$\pm$9.1e+00 & 1.73e-04$\pm$9.9e-05 & 3.12e-16$\pm$2.4e-16 & 3.52e+01 & 1.31e+02$\pm$1.9e+01 & 1.37e-03$\pm$2.0e-04 & 1.59e-15$\pm$2.3e-16\\
\hline
\end{tabular}
}
\end{table}

\begin{table}
\scalebox{0.6}{
\begin{tabular}{|cccccccc|}
\hline
\multicolumn{4}{|l}{1.0-2.0 keV Totals}  &  \multicolumn{4}{l|}{2.0-4.5 keV Totals} \\
(22) & (23) & (24) & (25) & (26) & (27) & (28) & (29)\\
DL   &  Counts  &  Count Rate  &  Flux & DL   &  Counts  &  Count Rate  &  Flux \\
    &   &  (s$^{-1}$)  &  (erg cm$^{-2}$ s$^{-1}$) &    &   &  (s$^{-1}$)  &  (erg cm$^{-2}$ s$^{-1}$)  \\
\hline
6.74e+02 & 7.73e+02$\pm$3.6e+01 & 7.70e-03$\pm$3.6e-04 & 8.32e-15$\pm$3.9e-16 & 3.04e+02 & 4.68e+02$\pm$3.0e+01 & 4.80e-03$\pm$3.1e-04 & 1.41e-14$\pm$9.1e-16\\
4.91e+00 & 3.62e+01$\pm$1.2e+01 & 3.84e-04$\pm$1.3e-04 & 2.35e-16$\pm$1.1e-16 & 2.20e+00 & 2.63e+01$\pm$1.1e+01 & 3.41e-04$\pm$1.4e-04 & 1.97e-16$\pm$2.0e-16\\
2.56e+01 & 5.15e+01$\pm$1.1e+01 & 1.15e-03$\pm$2.5e-04 & 8.93e-16$\pm$2.8e-16 & 9.91e-01 & 4.28e+00$\pm$6.5e+00 & 7.77e-05$\pm$1.4e-04 & 3.17e-16$\pm$5.2e-16\\
4.08e+00 & 4.11e+01$\pm$1.4e+01 & 3.54e-04$\pm$1.2e-04 & 1.43e-16$\pm$9.0e-17 & 6.50e+00 & 4.72e+01$\pm$1.4e+01 & 4.73e-04$\pm$1.4e-04 & 1.39e-15$\pm$3.9e-16\\
7.81e+00 & 7.27e+01$\pm$1.8e+01 & 7.72e-04$\pm$1.9e-04 & 8.43e-16$\pm$2.1e-16 & 1.95e-01 & 7.53e+00$\pm$1.2e+01 & 9.05e-05$\pm$1.4e-04 & 2.92e-16$\pm$4.0e-16\\
\hline
\end{tabular}
}
\end{table}

\begin{table}
\scalebox{0.55}{
\begin{tabular}{|cccccccccc|}
\hline
\multicolumn{5}{|l}{EPIC PN Parameters 0.2-0.5 keV}  &  \multicolumn{5}{l|}{EPIC PN Parameters 0.5-1.0 keV} \\
(30) & (31) & (32) & (33) & (34) & (35) & (36) & (37) & (38) & (39)\\
Expo  &  DL   &  Counts  &  Count Rate  &  Flux   &  Expo   &  DL   &  Counts  &  Count Rate  &  Flux \\
(ks)  &    &   &  (s$^{-1}$)  &  (erg cm$^{-2}$ s$^{-1}$)  &  (ks)  &    &   &  (s$^{-1}$)  &  (erg cm$^{-2}$ s$^{-1}$) \\
\hline
102.8 & 1.29e+01 & 4.08e+01$\pm$1.1e+01 & 3.97e-04$\pm$1.1e-04 & 1.43e-15$\pm$3.9e-16 & 102.7 & 1.59e+02 & 2.46e+02$\pm$2.2e+01 & 2.39e-03$\pm$2.1e-04 & 3.98e-15$\pm$3.5e-16\\
73.5 & 6.07e+00 & 1.96e+01$\pm$9.2e+00 & 2.66e-04$\pm$1.2e-04 & 9.60e-16$\pm$4.5e-16 & 73.5 & 5.97e-01 & 7.78e+00$\pm$7.8e+00 & 1.06e-04$\pm$1.1e-04 & 1.76e-16$\pm$1.8e-16\\
44.7 & 1.87e+00 & 5.21e+00$\pm$4.4e+00 & 1.17e-04$\pm$1.0e-04 & 4.21e-16$\pm$3.6e-16 & 44.6 & 4.15e+00 & 1.34e+01$\pm$7.0e+00 & 3.00e-04$\pm$1.6e-04 & 5.00e-16$\pm$2.6e-16\\
116.7 & 7.85e-01 & 1.16e+01$\pm$8.8e+00 & 9.91e-05$\pm$7.5e-05 & 3.58e-16$\pm$2.7e-16 & 116.6 & 0.00e+00 & 0.00e+00$\pm$5.9e+00 & 0.00e+00$\pm$5.0e-05 & 0.00e+00$\pm$8.4e-17\\
100.5 & 3.94e-01 & 5.19e+00$\pm$7.1e+00 & 5.16e-05$\pm$7.1e-05 & 1.86e-16$\pm$2.5e-16 & 100.4 & 2.78e+01 & 9.86e+01$\pm$1.7e+01 & 9.82e-04$\pm$1.7e-04 & 1.63e-15$\pm$2.8e-16\\
\hline
\end{tabular}
}
\end{table}

\begin{table}
\scalebox{0.55}{
\begin{tabular}{|cccccccccc|}
\hline
\multicolumn{5}{|l}{EPIC PN Parameters 1.0-2.0 keV}  &  \multicolumn{5}{l|}{EPIC PN Parameters 2.0-4.5 keV} \\
(40) & (41) & (42) & (43) & (44) & (45) & (46) & (47) & (48) & (49)\\
Expo  &  DL   &  Counts  &  Count Rate  &  Flux  &  Expo  &  DL   &  Counts  &  Count Rate  &  Flux \\
(ks)  &    &   &  (s$^{-1}$)  &  (erg cm$^{-2}$ s$^{-1}$)  &  (ks)  &    &   &  (s$^{-1}$)  &  (erg cm$^{-2}$ s$^{-1}$) \\
\hline
102.8 & 4.25e+02 & 4.82e+02$\pm$2.9e+01 & 4.69e-03$\pm$2.8e-04 & 8.56e-15$\pm$5.1e-16 & 99.8 & 1.38e+02 & 2.60e+02$\pm$2.4e+01 & 2.61e-03$\pm$2.4e-04 & 1.35e-14$\pm$1.2e-15\\
73.5 & 4.24e+00 & 1.53e+01$\pm$8.0e+00 & 2.08e-04$\pm$1.1e-04 & 3.79e-16$\pm$2.0e-16 & 72.2 & 4.64e+00 & 2.18e+01$\pm$9.7e+00 & 3.02e-04$\pm$1.3e-04 & 1.57e-15$\pm$7.0e-16\\
44.6 & 2.78e+01 & 5.15e+01$\pm$1.1e+01 & 1.15e-03$\pm$2.4e-04 & 2.10e-15$\pm$4.3e-16 & 44.2 & 6.18e-04 & 0.00e+00$\pm$5.3e+00 & 0.00e+00$\pm$1.2e-04 & 0.00e+00$\pm$6.2e-16\\
116.7 & 5.40e+00 & 3.46e+01$\pm$1.2e+01 & 2.96e-04$\pm$1.0e-04 & 5.41e-16$\pm$1.9e-16 & 114.2 & 3.09e+00 & 2.62e+01$\pm$1.1e+01 & 2.29e-04$\pm$9.9e-05 & 1.19e-15$\pm$5.2e-16\\
100.5 & 6.82e+00 & 4.70e+01$\pm$1.5e+01 & 4.68e-04$\pm$1.5e-04 & 8.53e-16$\pm$2.6e-16 & 97.5 & 2.01e-04 & 5.77e-01$\pm$9.6e+00 & 5.92e-06$\pm$9.8e-05 & 3.07e-17$\pm$5.1e-16\\
\hline
\end{tabular}
}
\end{table}

\begin{table}
\scalebox{0.55}{
\begin{tabular}{|cccccccccc|}
\hline
\multicolumn{5}{|l}{EPIC MOS1 Parameters 0.2-0.5 keV}  &  \multicolumn{5}{l|}{EPIC MOS1 Parameters 0.5-1.0 keV} \\
(50) & (51) & (52) & (53) & (54) & (55) & (56) & (57) & (58) & (59)\\
Expo  &  DL   &  Counts  &  Count Rate  &  Flux   &  Expo   &  DL   &  Counts  &  Count Rate  &  Flux \\
(ks)  &    &  &  (s$^{-1}$)  &  (erg cm$^{-2}$ s$^{-1}$)  &  (ks)  &    &   &  (s$^{-1}$)  &  (erg cm$^{-2}$ s$^{-1}$) \\
\hline
100.5 & 1.31e+00 & 3.97e+00$\pm$3.8e+00 & 3.95e-05$\pm$3.7e-05 & 7.88e-16$\pm$7.5e-16 & 100.4 & 5.77e+01 & 6.34e+01$\pm$1.0e+01 & 6.31e-04$\pm$1.0e-04 & 4.95e-15$\pm$8.0e-16\\
118.7 & 4.74e+00 & 9.03e+00$\pm$5.4e+00 & 7.61e-05$\pm$4.6e-05 & 1.52e-15$\pm$9.1e-16 & 118.6 & 2.98e-07 & 0.00e+00$\pm$4.0e+00 & 0.00e+00$\pm$3.3e-05 & 0.00e+00$\pm$2.6e-16\\
\nodata & \nodata & \nodata & \nodata & \nodata & \nodata & \nodata & \nodata & \nodata & \nodata\\
60.5 & 6.18e-04 & 0.00e+00$\pm$9.2e-01 & 0.00e+00$\pm$1.5e-05 & 0.00e+00$\pm$3.0e-16 & 60.5 & 6.18e-04 & 0.00e+00$\pm$1.0e+00 & 0.00e+00$\pm$1.7e-05 & 0.00e+00$\pm$1.4e-16\\
80.6 & 1.16e+00 & 4.80e+00$\pm$4.4e+00 & 5.95e-05$\pm$5.5e-05 & 1.19e-15$\pm$1.1e-15 & 80.6 & 6.58e+00 & 1.76e+01$\pm$6.9e+00 & 2.18e-04$\pm$8.5e-05 & 1.71e-15$\pm$6.7e-16\\
\hline
\end{tabular}
}
\end{table}

\begin{table}
\scalebox{0.55}{
\begin{tabular}{|cccccccccc|}
\hline
\multicolumn{5}{|l}{EPIC MOS1 Parameters 1.0-2.0 keV}  &  \multicolumn{5}{l|}{EPIC MOS1 Parameters 2.0-4.5 keV} \\
(60) & (61) & (62) & (63) & (64) & (65) & (66) & (67) & (68) & (69)\\
Expo  &  DL   &  Counts  &  Count Rate  &  Flux   &  Expo   &  DL   &  Counts  &  Count Rate  &  Flux \\
(ks)  &    &   &  (s$^{-1}$)  &  (erg cm$^{-2}$ s$^{-1}$)  &  (ks)  &    &   &  (s$^{-1}$)  &  (erg cm$^{-2}$ s$^{-1}$) \\
\hline
100.5 & 1.51e+02 & 1.70e+02$\pm$1.7e+01 & 1.69e-03$\pm$1.7e-04 & 9.08e-15$\pm$8.9e-16 & 98.2 & 1.07e+02 & 1.26e+02$\pm$1.5e+01 & 1.29e-03$\pm$1.5e-04 & 1.77e-14$\pm$2.1e-15\\
118.6 & 4.47e+00 & 1.96e+01$\pm$8.3e+00 & 1.65e-04$\pm$7.0e-05 & 8.85e-16$\pm$3.7e-16 & 116.7 & 2.98e-07 & 0.00e+00$\pm$1.9e+00 & 0.00e+00$\pm$1.6e-05 & 0.00e+00$\pm$2.2e-16\\
\nodata & \nodata & \nodata & \nodata & \nodata & \nodata & \nodata & \nodata & \nodata & \nodata\\
60.5 & 6.18e-04 & 0.00e+00$\pm$1.2e+00 & 0.00e+00$\pm$2.0e-05 & 0.00e+00$\pm$1.1e-16 & 60.2 & 3.89e+00 & 7.35e+00$\pm$4.3e+00 & 1.22e-04$\pm$7.2e-05 & 1.68e-15$\pm$9.9e-16\\
80.6 & 1.45e+00 & 1.02e+01$\pm$7.3e+00 & 1.27e-04$\pm$9.0e-05 & 6.81e-16$\pm$4.8e-16 & 78.5 & 4.13e-02 & 6.08e-01$\pm$5.9e+00 & 7.74e-06$\pm$7.5e-05 & 1.06e-16$\pm$1.0e-15\\
\hline
\end{tabular}
}
\end{table}

\begin{table}
\scalebox{0.55}{
\begin{tabular}{|cccccccccc|}
\hline
\multicolumn{5}{|l}{EPIC MOS2 Parameters 0.2-0.5 keV}  &  \multicolumn{5}{l|}{EPIC MOS2 Parameters 0.5-1.0 keV} \\
(70) & (71) & (72) & (73) & (74) & (75) & (76) & (77) & (78) & (79)\\
Expo  &  DL   &  Counts  &  Count Rate  &  Flux   &  Expo   &  DL   &  Counts  &  Count Rate  &  Flux \\
(ks)  &    &  &  (s$^{-1}$)  &  (erg cm$^{-2}$ s$^{-1}$)  &  (ks)  &    &   &  (s$^{-1}$)  &  (erg cm$^{-2}$ s$^{-1}$) \\
\hline
95.7 & 4.22e+00 & 1.05e+01$\pm$5.1e+00 & 1.10e-04$\pm$5.3e-05 & 2.21e-15$\pm$1.1e-15 & 95.7 & 6.68e+01 & 6.60e+01$\pm$1.0e+01 & 6.89e-04$\pm$1.1e-04 & 5.38e-15$\pm$8.5e-16\\
121.2 & 6.31e-02 & 6.63e-01$\pm$3.0e+00 & 5.47e-06$\pm$2.5e-05 & 1.10e-16$\pm$5.0e-16 & 121.2 & 5.79e-01 & 3.08e+00$\pm$3.9e+00 & 2.54e-05$\pm$3.2e-05 & 1.99e-16$\pm$2.5e-16\\
55.7 & 1.83e+00 & 3.00e+00$\pm$2.9e+00 & 5.39e-05$\pm$5.3e-05 & 1.08e-15$\pm$1.1e-15 & 55.7 & 6.18e-04 & 0.00e+00$\pm$2.8e+00 & 0.00e+00$\pm$5.1e-05 & 0.00e+00$\pm$4.0e-16\\
120.8 & 5.09e-02 & 7.28e-01$\pm$2.1e+00 & 6.03e-06$\pm$1.7e-05 & 1.21e-16$\pm$3.5e-16 & 120.8 & 1.35e+00 & 6.17e+00$\pm$4.7e+00 & 5.11e-05$\pm$3.9e-05 & 3.99e-16$\pm$3.1e-16\\
89.8 & 2.43e+00 & 5.20e+00$\pm$3.7e+00 & 5.79e-05$\pm$4.1e-05 & 1.16e-15$\pm$8.3e-16 & 89.7 & 7.23e+00 & 1.51e+01$\pm$5.7e+00 & 1.69e-04$\pm$6.3e-05 & 1.32e-15$\pm$4.9e-16\\
\hline
\end{tabular}
}
\end{table}

\begin{table}
\scalebox{0.55}{
\begin{tabular}{|cccccccccc|}
\hline
\multicolumn{5}{|l}{EPIC MOS2 Parameters 1.0-2.0 keV}  &  \multicolumn{5}{l|}{EPIC MOS2 Parameters 2.0-4.5 keV} \\
(80) & (81) & (82) & (83) & (84) & (85) & (86) & (87) & (88) & (89)\\
Expo  &  DL   &  Counts  &  Count Rate  &  Flux  &  Expo  &  DL   &  Counts  &  Count Rate  &  Flux \\
(ks)  &    &   &  (s$^{-1}$)  &  (erg cm$^{-2}$ s$^{-1}$)  &  (ks)  &    &   &  (s$^{-1}$)  &  (erg cm$^{-2}$ s$^{-1}$) \\
\hline
95.7 & 1.10e+02 & 1.21e+02$\pm$1.4e+01 & 1.26e-03$\pm$1.5e-04 & 6.76e-15$\pm$8.0e-16 & 93.7 & 6.91e+01 & 8.18e+01$\pm$1.2e+01 & 8.73e-04$\pm$1.3e-04 & 1.20e-14$\pm$1.8e-15\\
121.2 & 7.25e-02 & 1.36e+00$\pm$3.2e+00 & 1.12e-05$\pm$2.7e-05 & 6.01e-17$\pm$1.4e-16 & 119.4 & 5.24e-01 & 4.50e+00$\pm$4.5e+00 & 3.77e-05$\pm$3.8e-05 & 5.16e-16$\pm$5.2e-16\\
55.7 & 6.18e-04 & 0.00e+00$\pm$3.8e+00 & 0.00e+00$\pm$6.9e-05 & 0.00e+00$\pm$3.7e-16 & 55.1 & 1.84e+00 & 4.28e+00$\pm$3.8e+00 & 7.78e-05$\pm$7.0e-05 & 1.06e-15$\pm$9.5e-16\\
120.8 & 1.21e+00 & 6.51e+00$\pm$7.3e+00 & 5.39e-05$\pm$6.1e-05 & 2.89e-16$\pm$3.3e-16 & 119.2 & 4.04e+00 & 1.37e+01$\pm$6.4e+00 & 1.15e-04$\pm$5.4e-05 & 1.57e-15$\pm$7.3e-16\\
89.7 & 3.77e+00 & 1.55e+01$\pm$7.2e+00 & 1.72e-04$\pm$8.0e-05 & 9.23e-16$\pm$4.3e-16 & 87.7 & 1.76e+00 & 6.34e+00$\pm$4.9e+00 & 7.23e-05$\pm$5.6e-05 & 9.89e-16$\pm$7.6e-16\\
\hline
\end{tabular}
}
\end{table}

\begin{table}
\scalebox{0.55}{
\begin{tabular}{|cccc|}
\hline
(90) & (91) & (92) & (93)\\
HR1  & HR2  &  HR$_{1C}$  & HR$_{2C}$\\
\hline
 0.09 & 0.2 & 0.74 & 0.35\\
-0.48 & -0.03 & -0.46 & 0.54\\
 0.03 & -0.28 & 0.24 & 0.59\\
-0.03 & 0.72 & -0.33 & 0.74\\
-0.35 & -0.18 & 0.79 & -0.29\\
\hline
\end{tabular}
}
\end{table}


\clearpage

\begin{table}
\caption{Possible Transients}
\begin{center}
\begin{tabular}{ccccccc}
\hline
\hline
M06 ID & T11 ID & Source (Herein) & Peak (erg~cm$^{-2}$~s$^{-1}$) & Limit (erg~cm$^{-2}$~s$^{-1}$) & Ratio & Peak S/N\\
\hline
 41 & \nodata & \nodata & 6.56e-14 & $<$8.95e-16 & 73.30 & 18.25 \\ 
96 & \nodata & \nodata & 2.76e-13 & $<$1.38e-15 & 200.12 & 20.89 \\ 
134 & \nodata & \nodata & 1.59e-13 & $<$3.12e-15 & 50.90 & 15.00 \\ 
149 & \nodata & \nodata & 1.21e-14 & $<$1.04e-15 & 11.68 & 5.08 \\
180 & \nodata & \nodata & 3.63e-14 & $<$1.66e-15 & 21.88 & 6.86 \\ 
207 & \nodata & \nodata & 1.10e-13 & $<$1.19e-15 & 92.79 & 14.44 \\ 
246 & \nodata & \nodata & 1.81e-14 & $<$1.53e-15 & 11.82 & 11.71 \\
296 & 634? & \nodata & 5.29e-14 & $<$1.32e-15 & 40.07 & 8.45 \\ 
\nodata & 13 & \nodata & 1.10e-14 & $<$8.50e-16 & 12.98 & 14.15\\ 
\nodata & 26 & \nodata & 9.70e-15 & $<$9.16e-16 & 10.59 & 12.18 \\ 
\nodata & 233 & \nodata & 1.41e-14 & $<$1.24e-15 & 11.37 & 20.05 \\
\nodata & 283 & \nodata & 1.96e-14 & $<$1.43e-15 & 13.73 & 25.01 \\
\nodata & \nodata & 712 & 4.83e-14 & $<$3.00e-16 & 161.10 & 49.80\\ 
113 & \nodata & 426 & 5.39e-14 & $<$3.00e-16& 179.81 & 26.80 \\ 
228 & \nodata & 859 & 1.03e-14 &$<$ 3.00e-16 & 34.20 & 14.22 \\
285 & \nodata & 1032 & 7.07e-15 & $<$3.00e-16 & 23.58 & 11.72 \\
236 & \nodata & 882 & 6.98e-15 & $<$3.00e-16 & 23.26 & 12.39 \\
\nodata & \nodata & 633\tablenotemark{1} & 6.41e-15 & $<$3.00e-16 & 21.37 & 12.83 \\ 
\nodata & \nodata & 551 & 1.30e-14 &$<$ 3.00e-16 & 43.30 & 14.38 \\
\nodata & \nodata & 556 & 1.91e-14 & $<$3.00e-16 & 63.82 & 13.80 \\
255 & \nodata & 940 & 1.13e-14 & $<$3.00e-16 & 37.83 & 13.65 \\
312 & \nodata & 1141 & 8.23e-15 &$<$ 3.00e-16 & 27.45 & 10.80 \\
\nodata & \nodata & 887\tablenotemark{1} & 5.11e-15 & $<$3.00e-16 & 17.03 & 10.31 \\ 
\nodata & \nodata & 916 & 1.26e-14 & $<$3.00e-16 & 41.89 & 12.15 \\
\nodata & \nodata & 17 & 8.49e-15 & $<$3.00e-16 & 28.30 & 10.99 \\
206 & \nodata & 784 & 3.45e-15 & $<$3.00e-16 & 11.50 & 8.06 \\
\nodata & \nodata & 1043 & 4.64e-15 & $<$3.00e-16 & 15.47 & 8.62 \\
\nodata & \nodata & 637 & 4.34e-15 & $<$3.00e-16 & 14.46 & 8.99 \\
\nodata & \nodata & 748 & 8.25e-15 & $<$3.00e-16 & 27.50 & 9.61 \\
\nodata & \nodata & 475 & 3.78e-15 & $<$3.00e-16 & 12.60 & 8.84 \\
\nodata & \nodata & 55 & 8.29e-15 & $<$3.00e-16 & 27.65 & 8.37 \\
\nodata & \nodata & 514 & 8.21e-15 & $<$3.00e-16 & 27.38 & 8.76 \\
\nodata & \nodata & 477 & 3.95e-15 & $<$3.00e-16 & 13.17 & 9.05 \\
\hline
\tablenotetext{1}{Foreground Star}
\label{transients}
\end{tabular}
\end{center}
\end{table}

\clearpage


\begin{table}
\caption{Variability of matched T11 sources: Full version available in machine-readable format only.}
\begin{center}
\begin{tabular}{cccccc}

\hline\hline
Source Number & XMM Flux & T11 Flux & T11 Revised Flux &  Sigma & Sigma Revised\\
\hline
177 & 2.21e-15 & 3.64e-15 & 6.43e-15 & 3.46 & 3.02 \\
189 & 2.07e-15 & 1.28e-15 & 2.26e-15 & 0.26 & 0.23 \\
198 & 1.03e-14 & 4.93e-15 & 8.69e-15 & 1.22 & 0.86 \\
200 & 3.38e-15 & 2.84e-15 & 5.01e-15 & 1.76 & 1.48 \\
202 & 3.17e-15 & 2.16e-15 & 3.81e-15 & 0.85 & 0.71 \\
203 & 1.24e-14 & 5.79e-15 & 1.02e-14 & 2.00 & 1.14 \\
217 & 6.25e-16 & 6.75e-16 & 1.19e-15 & 1.19 & 1.15 \\
222 & 3.78e-15 & 1.33e-15 & 2.34e-15 & 1.50 & 1.36 \\
234 & 2.25e-15 & 2.76e-14 & 4.87e-14 & 25.47 & 8.92 \\
252 & 1.18e-15 & 7.39e-16 & 1.30e-15 & 0.32 & 0.29 \\
\nodata & \nodata & \nodata & \nodata & \nodata & \nodata \\
\hline
\label{matchedt11}
\end{tabular}
\end{center}
\end{table}

\clearpage

\begin{table}
\begin{center}
\caption{Short Term Variability}
\begin{tabular}{lrrccl}
\hline\hline\noalign{\smallskip}
 OBSID      &  Source  &  P04    & M06  & T11 &    Comment\tablenotemark{1}                    \\
\hline\hline\noalign{\smallskip}
 0650510201 &  521   &  171       & 150        & 225         & X-7 eclipsing HMXB                \\
 0650510201 &  712  &    -       &   -        &  -          & 285.4 s pulsar \citep{trudolyubov2013} \\
 0650510101 &  712  &    -       &   -        &  -          & 285.4 s pulsar \citep{trudolyubov2013} \\
 0650510201 &  763  &  240       & 203        & 409         & fg star V=19.2                    \\
 0650510601 &  763  &  240       & 203        & 409         & fg star V=19.2                    \\
 0672190301 &  1166  &  374       & 320        &  -          & fg star G4, V=9.6, high PPM       \\
 0672190301 &  1271  &  406       & 348        &  -          & fg star F5 \citep{hatzidimitriou2006} \\
 0650510301 & 933  &  297       & 253        &  -          & fg star G8 \citep{hatzidimitriou2006} \\
 0650510301 &  1087  &  346       & 299        &  -          & $<$fg star$>$ V=15.7              \\
 0650510601 &  550  &  182       & 156        & 242         & fg star A5 V=8.1                  \\
 0650510701 &  303  &   77       &  71        &  -          & fg star F5 \citep{hatzidimitriou2006}\\
 0606370401 &  128  &    -       &   -        &  -          & transient? (only 1 detection      \\
            &      &                    &            &             & in monitoring of [PMH2004]47)     \\
            &      &                 &            &             & variable in M33? \citep{hartman2006}  \\
 0650510701 &  213   &     -       &   -        &  -          & variable in M33? \citep{hartman2006}  \\
 0650510301 &  995   &  320        & 273        &  -          & $<$fg star$>$ V=18.3              \\
 0672190301 &  1095   &  349        & 301        & 640         & $<$fg star$>$ V=18.1              \\
 0672190301 &  1200   &  -          & -          & -           & no previous X-ray detection       \\
            &      &                &            &             & $<$fg star$>$ V=16.0              \\ 
 0650510701 &  667   &   -          & -          & -           & no previous X-ray detection       \\
            &      &                 &            &             & $<$fg star$>$ V=16.9              \\ 
\hline\hline\noalign{\smallskip}
\end{tabular}
\tablenotetext{1}{$<>$ symbols denote a preliminary classification.  Absence of
  these brackets denotes firm identification;  fg is short for foreground.}
\end{center}
\label{variables}
\end{table}


\clearpage
\newpage

\end{document}